\documentclass[12pt]{emulateapj}
\usepackage{natbib,xcolor}
\usepackage[caption=false]{subfig}    
\shorttitle{Stellar Mass Fractions in $z\sim1$ \madcows\ Clusters}
\shortauthors{Decker et al.}

\newcommand{\eg}{e.g.,}

\newcommand{\ie}{i.e.,}
\newcommand{\spitzer}{{\it Spitzer}}

\newcommand{\msol}{M_\odot}

\newcommand{\um}{$\mathrm{\mu m}$}
\newcommand{\madcows}{MaDCoWS}
\newcommand{\mfh}{$M_{500}$}
\newcommand{\rfh}{$r_{500}$}
\newcommand{\teneft}{\times10^{14}}
\newcommand{\mstar}{m^*}

\newcommand{\cho}{$3.6~\mathrm{\mu m}$}
\newcommand{\cht}{$4.5~\mathrm{\mu m}$}
\newcommand{\fstar}{$f_\star$}
\newcommand{\fgas}{$f_{\mathrm{gas}}$}
\newcommand{\rank}{third}
\newcommand{\nSPT}{33}
\newcommand{\Nmadcows}{twelve}
\newcommand{\newSZ}{seven}
\newcommand{\nopt}{five}
\newcommand{\irmean}{$f_\star = 0.015\pm0.005$}

\newcommand{\irnormmean}{$f_\star/G13 = 1.16\pm0.12$}
\newcommand{\sznormmean}{$f_\star/G13 = 0.88\pm0.09$}
\newcommand{\normsig}{$1.9~\sigma$}
\newcommand{\fitms}{19.41}

\begin{document}
\title{The Massive and Distant Clusters of WISE Survey VI: Stellar
  Mass Fractions of a Sample of High-Redshift Infrared-selected
  Clusters}
\author{Bandon Decker\altaffilmark{1},
Mark Brodwin\altaffilmark{1},
Zubair Abdulla\altaffilmark{2,3},
Anthony H. Gonzalez\altaffilmark{4},
Daniel P. Marrone\altaffilmark{5},
Christine O'Donnell\altaffilmark{5},
S. A. Stanford\altaffilmark{6},
Dominika Wylezalek\altaffilmark{7},
John E. Carlstrom\altaffilmark{2,3},
Peter R. M. Eisenhardt\altaffilmark{8},
Adam Mantz\altaffilmark{9,10},
Wenli Mo\altaffilmark{4},
Emily Moravec\altaffilmark{4},
Daniel Stern\altaffilmark{8},
Greg Aldering\altaffilmark{11},
Matthew L. N. Ashby\altaffilmark{12},
Kyle Boone\altaffilmark{11,13},
Brian Hayden\altaffilmark{14,11},
Nikhel Gupta\altaffilmark{15},
Michael A. McDonald\altaffilmark{16}
}
\altaffiltext{1}{Department of Physics and Astronomy, University of Missouri, 5110 Rockhill Road, Kansas City, MO 64110, USA}
\altaffiltext{2}{Kavli Institute for Cosmological Physics, University of Chicago, 5640 South Ellis Avenue, Chicago, IL 60637, USA}
\altaffiltext{3}{Department of Astronomy and Astrophysics, University of Chicago, 5640 South Ellis Avenue, Chicago, IL 60637, USA}
\altaffiltext{4}{Department of Astronomy, University of Florida, 211 Bryant Space Center, Gainesville, FL 32611, USA}
\altaffiltext{5}{Steward Observatory, University of Arizona, 933 North Cherry Avenue, Tucson, AZ 85721, USA}
\altaffiltext{6}{Department of Physics, University of California, One Shields Avenue, Davis, CA 95616, USA}
\altaffiltext{7}{European Southern Observatory, Karl Schwarzschild Stra\ss e 2, D-85748, Garching bei M\"{u}nchen, Germany}
\altaffiltext{8}{Jet Propulsion Laboratory, California Institute of Technology, Pasadena, CA 91109, USA}
\altaffiltext{9}{Kavli Institute for Particle Astrophysics and Cosmology, Stanford University, 382 Via Pueblo Mall, Stanford, CA 94305-4060, USA}
\altaffiltext{10}{Department of Physics, Stanford University, 382 Via Pueblo Mall, Stanford, CA 94305-4060, USA}
\altaffiltext{11}{Lawrence Berkeley National Laboratory, 1 Cyclotron Road, MS 50B-4206, Berkeley, CA 94720, USA}
\altaffiltext{12}{Harvard-Smithsonian Center for Astrophysics, 60 Garden Street, Cambridge, MA 02138, USA}
\altaffiltext{13}{Department of Physics, University of California, 366 LeConte Hall MC 7300, Berkeley, CA 94720, USA}
\altaffiltext{14}{Space Telescope Science Institute, 3700 San Martin Drive, Baltimore, MD 21218, USA}
\altaffiltext{15}{School of Physics, University of Melbourne, Grattan Street, Parkville, VIC 3010, Australia}
\altaffiltext{16}{Kavli Institute for Astrophysics and Space
Research, Massachusetts Institute of Technology, 77 Massachusetts
Avenue, Cambridge, MA 02139, USA}

\begin{abstract}
We present measurements of the stellar mass fractions (\fstar) for a
sample of high-redshift ($0.93 \le z \le 1.32$) infrared-selected
galaxy clusters from the Massive and Distant Clusters of WISE Survey
(\madcows) and compare them to the stellar mass fractions of
Sunyaev-Zel'dovich (SZ) effect-selected clusters in a similar mass and
redshift range from the South Pole Telescope (SPT)-SZ Survey. We do not find a significant difference in mean
\fstar\ between the two selection methods, though we do find an
unexpectedly large range in \fstar\ for the SZ-selected
clusters. In addition, we measure the luminosity function of the
\madcows\ clusters and find $\mstar= \fitms\pm0.07$, similar to other
studies of clusters at or near our redshift range. Finally, we present
SZ detections and masses for seven \madcows\ clusters and new
spectroscopic redshifts for five \madcows\ clusters. One of
these new clusters, MOO\ J1521+0452 at $z=1.31$, is the most distant
\madcows\ cluster confirmed to date. 
\end{abstract}

\section{Introduction}
\label{Sec: Intro}
Galaxy clusters are the largest gravitationally-bound objects in the
universe and a thorough knowledge of their composition, history and
evolution is important for both cosmological abundance analyses and
galaxy formation/evolution studies in the richest environments
\citep[\eg][]{Allen+11,KB12}. It has been found in simulations
\citep[\eg][]{Ettori+06,Conroy+07} and suggested observationally
\citep[\eg][]{LMS03} that the fraction of a cluster's total mass that
is in stars, \fstar, is related to the star formation history of that
cluster. It follows that measuring \fstar\ and \fgas, the fraction of
mass in the intracluster medium (ICM), in clusters covering a
range of masses and ages can constrain the growth and evolutionary
history of clusters and the galaxies therein. A proper account of the
total stellar mass of a cluster is also a necessary component of
calculating the total baryon fraction in a cluster. The cluster
baryon fraction is expected to be close to the total baryon fraction
of the universe \citep{White+93}, but previous studies
\citep[\eg][]{GZZ07,GZZ13,Lin+12} have found somewhat lower
fractions. The size of this discrepancy and its relation to the total
mass of the cluster is important cosmologically and can also provide
clues to the baryon physics in clusters \citep{He+05}. Coupled with
studies that show a cessation of star formation in the cores of large
clusters since high redshifts \citep[\eg][]{Brodwin+13}, such
measurements can shed light on the feedback processes involved in the
partition of baryons into stars and gas in clusters.

Several studies have previously looked at the stellar mass fraction of
clusters and generally find a trend of decreasing \fstar\ with
increasing halo mass. However, with the exception of
\citet{vanderBurg+14}, all these studies were at $z\le0.6$
\citep{LMS03,GZZ07,Andreon10,Zhang+11,Lin+12,GZZ13} and/or used
samples that selected clusters entirely on the strength of the signal
from the ICM, either from X-ray observations or from the
Sunyaev-Zel'dovich \citep[SZ,][]{SZ70,SZ72} decrement
\citep{Giodini+09,Hilton+13,Chiu+16,Chiu+18}. It is possible, for both
SZ- and X-ray-selected samples, that selecting on an observable
related to the ICM pressure or X-ray luminosity (approximately ICM
density squared) could produce a sample with a bias toward a higher
gas mass fraction, presumably at the expense of \fstar\ (assuming a
constant baryon fraction at fixed mass). Such a bias may also prevent
the scatter in \fgas\ from being fairly measured, though the measured
scatter in \fstar\ should be less affected, as the cluster
selection does not have any intrinsic bias toward or against stellar
mass. 

To explore these issues, we use high-redshift infrared-selected
clusters from the Massive and Distant Clusters of {\it WISE} Survey
\citep[\madcows,][]{Gettings+12,Stanford+14,Brodwin+15,Gonzalez+15,Mo+18,Gonzalez+18}. \madcows\
uses the {\it Wide-field Infrared Survey Explorer} \citep[{\it
WISE},][]{WISE} All{\it WISE} data release \citep{Cutri13} and
PanSTARRS \citep{PanSTARRS} optical data to identify overdensities of
galaxies at $z\sim1$ across nearly the entire extragalactic sky. It
therefore can provide a greater mass range at high-redshift than SZ
surveys because it simultaneously has the area to find the rarest,
most massive objects at high redshifts---such as MOO\ J1142+1527
(\mfh\ = $5.36~\teneft~\msol$, $z=1.19$) reported in
\citet{Gonzalez+15} and MOO\ J1521+0452 (\mfh\ = $3.59~\teneft~\msol$,
$z=1.31$) described herein---and the sensitivity to detect clusters
to the same or lower mass limit of current SZ surveys.

In this work we use SZ observations and follow-up {\it Spitzer
Space Telescope} data on \Nmadcows\ \madcows\ clusters to calculate
\fstar\ for this high-redshift, infrared-selected sample. We also
analyze a comparable sample of SZ-selected clusters from the
South Pole Telescope (SPT)-SZ survey \citep{Bleem+15} using the same
methodology and compare these to the same quantities measured for our
infrared-selected \madcows\ clusters. Because the SPT sample is
SZ-selected, it fairly measures the average value and scatter
in \fstar.

The cluster samples and data we use are described in
\textsection{\ref{Sec: sample}} and the analysis thereof is described
in \textsection{\ref{Sec: analysis}}. We present the results of our
\fstar\ measurements in \textsection{\ref{Sec: results}} and discuss
them in \textsection{\ref{Sec: discussion}}. Our conclusions are in
\textsection{\ref{Sec: conclusion}}. Throughout we use AB magnitudes
and a concordance $\Lambda$CDM cosmology of $\Omega_m = 0.3$,
$\Omega_\Lambda = 0.7$ and $H_0 = 70~\mathrm{km~s^{-1} Mpc^{-1}}$. We
define \rfh\ as the radius inside which a cluster has an average
density 500 times the critical density of the universe and \mfh\ as
the mass inside that radius.

\section{Cluster Sample and Data}
\label{Sec: sample} 
For our infrared-selected sample, we use twelve \madcows\ clusters
with halo masses calculated from SZ detections from the
Combined Array for Research in Millimeter-wave Astronomy (CARMA). SZ
observations of four of these (MOO\ J0319-0025, MOO\ J1014+0038, MOO\
J1155+3901 and MOO\ J1514+1346) are described in \citet{Brodwin+15}. A
fifth, MOO\ J1142+1527, the most massive cluster yet found by
any method at $z\ge1.15$, is reported in \citet{Gonzalez+15}. 
Here we report new SZ detections for the other \newSZ\ clusters,
along with total masses and radii determined from those data as well
as new masses and radii of the previously-reported clusters derived
from an updated reduction of the CARMA data, described in
\textsection{\ref{Ssec: mass}}. All twelve clusters have imaging with
the Infrared Array Camera \citep[IRAC,][]{IRAC} on \spitzer, which
enables us to determine the stellar mass of the clusters as described
in \textsection\ref{Ssec: stellar mass}.

\begin{figure}[bthp] 
\center{\includegraphics[width=8.5cm]{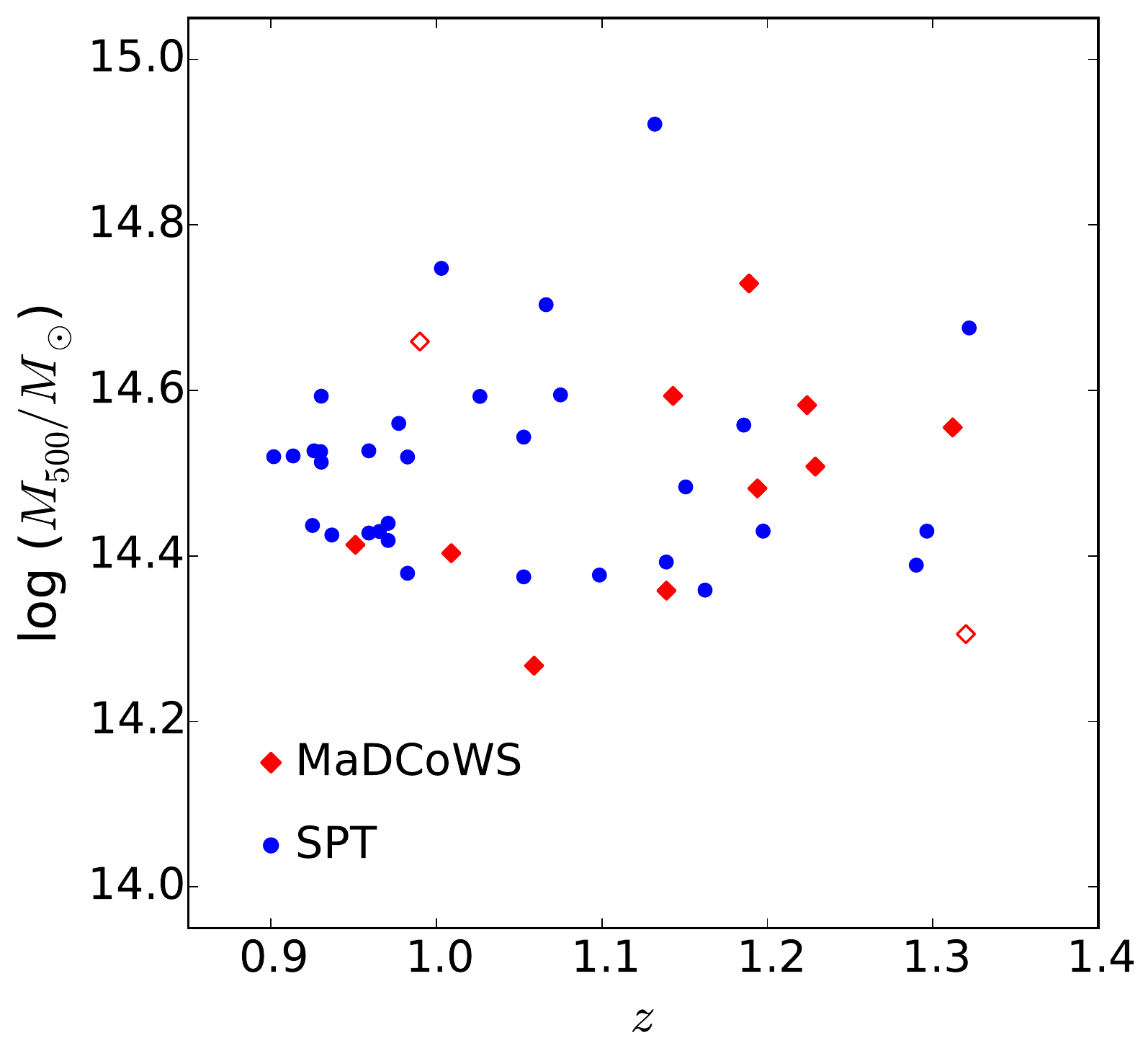}}
\caption{Plot showing the distribution of the \madcows\ clusters
  (red diamonds) and the comparison SPT clusters (blue circles) in
  mass and redshift. The open diamonds denote the \madcows\ clusters
  in this analysis that currently lack spectroscopic redshifts.}
\label{Fig: mass v z} 
\end{figure}

The SZ-selected clusters we use for comparison are drawn from
the SPT-SZ survey described in \citet{Bleem+15}. To ensure we are
making a fair comparison between the infrared- and SZ-selected
samples, we only use the \nSPT\ SPT clusters that lie in a similar
range of mass and redshift as the \madcows\ clusters, specifically
$0.9 < z < 1.35$ and \mfh\ $< 1\times10^{15}~\msol$, and for which
comparable IRAC data exist. We do not impose a lower limit on the mass
for the SPT subsample as the SPT-SZ catalog has a higher mass
threshold than \madcows\ at these redshifts. A plot of mass versus
redshift for both samples is shown in Figure \ref{Fig: mass v z}.

\subsection{CARMA Data}
\label{Ssec: Carma}
Before its closure in early 2015, CARMA was a heterogenous 23-element
interferometer with six 10.4 m antennae, nine 6.1 m antennae and eight
3.5 m antennae. All of the antennae were equipped with 30 and 90 GHz
receivers and the 10.4 and 6.1 m antennae had additional 230 GHz
receivers. CARMA had two correlators, a wide-band (WB) and
spectral-line (SL) correlator, and the 3.5 m antennae could operate as
an independent array (CARMA-8 mode) or alongside the other 15 antennae
in CARMA-23 mode. In its most compact `E' configuration, the shortest
CARMA baselines provided an appropriate beam size for
SZ observations while the longer baselines enabled point source
identification and subtraction.

\begin{figure*}[bthp]
\center{\includegraphics[width=18cm]{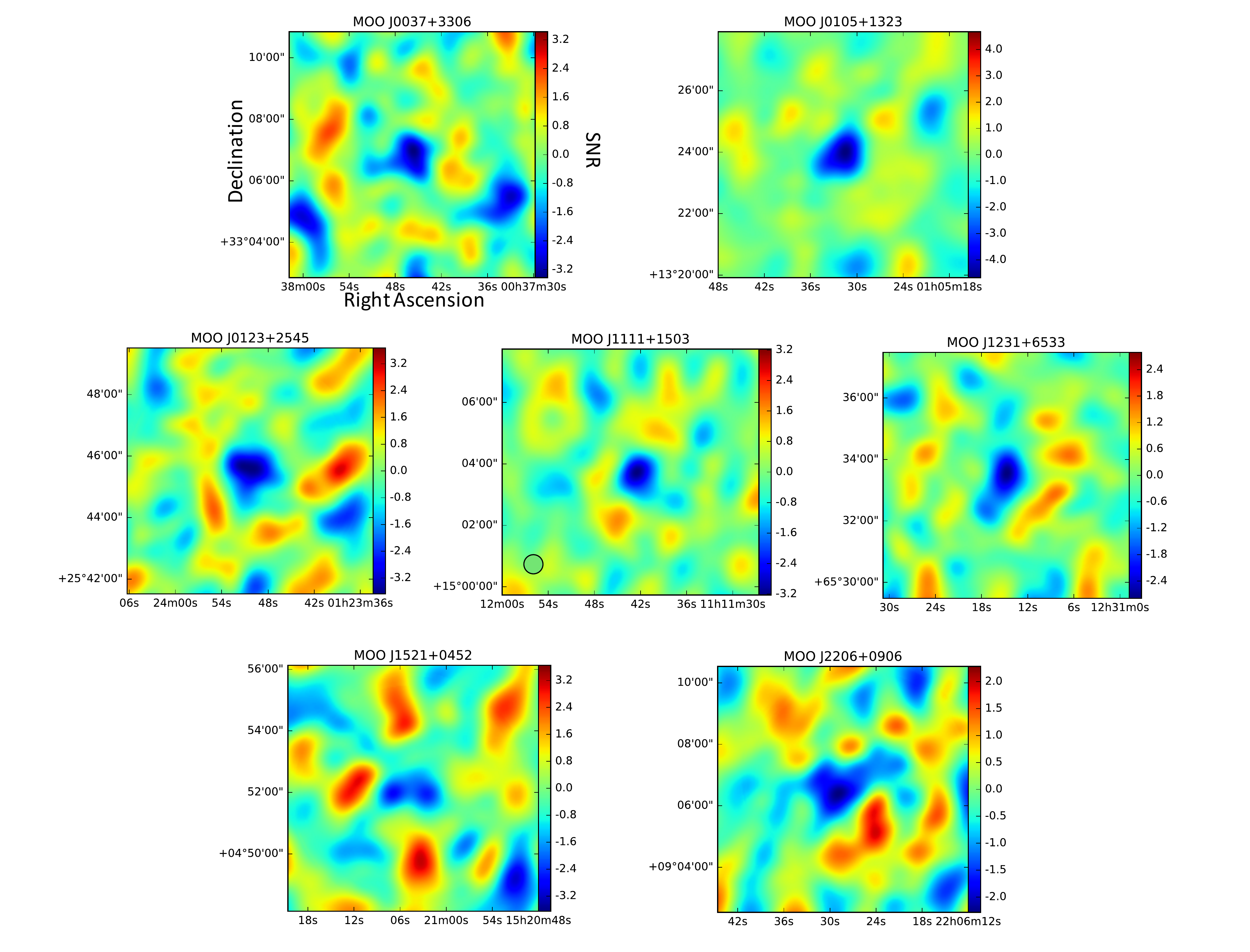}}
\caption{CARMA maps of the \newSZ\ new \madcows\ clusters presented
here. Each map is $8\arcmin \times 8\arcmin$, centered on the centroid of
the SZ decrement and in units of signal-to-noise. Emissive
point sources have been subtracted out of all the maps and they
have all been CLEANed around the decrement. A representative beam
pattern is shown in the lower left-hand corner of the map of MOO\
J1111+1503.}
\label{Fig: New clusters}
\end{figure*}

The CARMA data for the \newSZ\ new clusters were taken in the summer and
autumn of 2014 and the observation dates of all \Nmadcows\ of our
\madcows\ clusters, as well as the on-source observation times excluding
observations of the gain and flux calibrators, are given in Table
\ref{Table: Carma observations}. 
Point source-subtracted SZ maps of the \newSZ\ clusters newly reported
here are shown in Figure \ref{Fig: New clusters}. The maps are in
units of signal-to-noise with negative signal to denote the SZ effect
being a decrement at 30 GHz. A 4k$\lambda$ taper was applied to the
{\it uv} data to produce an illustrative beam size and the maps were
CLEANed \citep{CLEAN} in a box $4\arcmin$ on a side and centered on
the SZ centroid.

\begin{deluxetable*}{lllccclccl}
  \tabletypesize{\normalsize} \tablecaption{Summary of CARMA
    observations and properties of \madcows\ clusters\label{Table: Carma observations}} \tablewidth{0pt} \tablehead{
    \colhead{Cluster ID} & \colhead{RA} & \colhead{Dec.} &
    \colhead{UT Dates} &     \colhead{Exp.~Time\tablenotemark{a}} &
    \colhead{S/N} &
    \colhead{$z$} & \colhead{$Y_{500}$} & \colhead{\rfh} & \colhead{\mfh} \\
    \colhead{} & \colhead{(J2000)} & \colhead{(J2000)} &
    \colhead{}& \colhead{(hr)} & \colhead{($\sigma$)} & \colhead{} &
    \colhead{($10^{-5}~\mathrm{Mpc}^{2}$)} & \colhead{(Mpc)} & \colhead{($10^{14}~\msol$)} } \startdata
  MOO\ J0037$+$3306 & 00:37:45.8 & $+$33:06:51 & 2014 Sep 12,27-28 & 6.0 &
  3.3 & 1.133 & $1.78^{+0.87}_{-0.73}$ & $0.62^{+0.05}_{-0.06}$ & $2.26^{+0.62}_{-0.61}$ \\
  MOO\ J0105$+$1323\tablenotemark{e} & 01:05:31.5 & $+$13:23:55 & 2014 Jul 6; Oct 11 &
  7.3 & 8.1 & 1.143 & $1.49^{+0.91}_{-0.80}$ & $0.72 \pm 0.03$ & $3.92^{+0.46}_{-0.44}$ \\
  MOO\ J0123$+$2545 & 01:23:50.3 & $+$25:45:31 & 2014 Sep 27 & 1.9 &
  4.4 & 1.224 & $4.47^{+1.76}_{-1.43}$ & $0.70 \pm 0.05$ & $3.86^{+0.85}_{-0.79}$ \\
  MOO\ J0319$-$0025\tablenotemark{b} & 03:19:24.4 & $-$00:25:21 & 2013
  Sep 30 & 1.0 & 5.7 & 1.194 & $2.97^{+0.75}_{-0.78}$ & $0.65^{+0.03}_{-0.04}$ & $2.97^{+0.75}_{-0.78}$ \\
  MOO\ J1014$+$0038\tablenotemark{b} & 10:14:08.4 & $+$00:38:26 &
  2013 Oct 6-7 & 2.2 & 8.0 & 1.231 & $3.34^{+0.64}_{-0.52}$ & $0.66 \pm 0.02$ & $3.22^{+0.36}_{-0.31}$ \\
  MOO\ J1111$+$1503 & 11:11:42.6 & $+$15:03:44 & 2014 Jul 23,25 & 4.4
  & 5.0 & 1.32\tablenotemark{d} & $1.58^{+0.41}_{-0.37}$ & $0.54 \pm 0.04$ & $2.02^{+0.30}_{-0.30}$ \\
  MOO\ J1142$+$1527\tablenotemark{c} & 11:42:45.1 & $+$15:27:05 & 2014
  Jul 3 & 5.0 & 10.4 & 1.189 & $7.70^{+1.36}_{-1.17}$ & $0.79 \pm 0.03$ & $5.36^{+0.55}_{-0.50}$ \\
  MOO\ J1155$+$3901\tablenotemark{b} & 11:55:45.6 & $+$39:01:15 & 2012
  May 11-12 & 7.2 & 2.9 & 1.009 & $2.05^{+0.72}_{-0.65}$ & $0.66^{+0.04}_{-0.05}$ & $2.53^{+0.50}_{-0.51}$ \\
  MOO\ J1231$+$6533 & 12:31:14.8 & $+$65:33:29 & 2014 Sep 7-8 & 1.5 &
  4.3 & 0.99\tablenotemark{d} & $5.76^{+2.83}_{-1.80}$ & $0.81^{+0.07}_{-0.06}$ & $4.69^{+1.27}_{-0.94}$ \\
  MOO\ J1514$+$1346\tablenotemark{b,e} & 15:14:42.7 & $+$13:46:31 & 2013
  Jun 1,3,5-7,9,11 & 8.4 & 2.8 & 1.059 & $1.91^{+0.73}_{-0.97}$ & $0.63^{+0.04}_{-0.08}$ & $2.39^{+0.51}_{-0.83}$ \\
  MOO\ J1521$+$0452 & 15:21:04.6 & $+$04:52:08 & 2014 Sep 23 & 2.5 &
  2.7 & 1.312 & $4.13^{+2.14}_{-1.61}$ & $0.66 \pm 0.07$ & $3.59^{+1.02}_{-0.92}$ \\
  MOO\ J2206$+$0906\tablenotemark{e} & 22:06:28.6 & $+$09:06:32 & 2014 Jul 5,8 & 5.7 &
  3.1 & 0.926 & $2.58^{+1.30}_{-0.92}$ & $0.71 \pm 0.06$ & $2.95^{+0.82}_{-0.68}$  
\enddata
\tablenotetext{a}{On-source, unflagged.}
\tablenotetext{b}{\citet{Brodwin+15}, with a mass and radius
  re-calculated from an improved CARMA reduction pipeline.}
\tablenotetext{c}{\citet{Gonzalez+15}, with a mass and radius
  re-calculated from an improved CARMA reduction pipeline and using a slightly different cosmology.}
\tablenotetext{d}{Photometric redshift estimated from IRAC \cho\ and 
  \cht\ images, with error $\sim0.07$.}
\tablenotetext{e}{Identified as a merging cluster from follow-up {\it
    Chandra} imaging \citep[see ][]{Gonzalez+18}.}
\end{deluxetable*}

\subsection{Spitzer Data}
\label{Ssec: Spitzer}
Eight of the \madcows\ clusters were observed in \spitzer\ Cycle 9
(Program ID 90177; PI Gonzalez) and have
$6\mathrm{\times}30~\mathrm{s}$ depth in the IRAC \cho\ and \cht\
channels, while the remaining four were observed to the same depth as
part of a Cycle 11-12 snapshot program (PID 11080; PI Gonzalez). This
depth allows us to detect objects down to one magnitude fainter than
the characteristic magnitude ($\mstar$) on all of our clusters with
high ($>70$\%) completeness. The SPT clusters were observed with
\spitzer\ over four Cycles (PID 60099, 70053, 80012, 10101; PI
Brodwin) to a depth of $8\mathrm{\times}100~\mathrm{s}$ in \cho\ and
$6\mathrm{\times}30~\mathrm{s}$ in \cht.

\subsection{Optical Data}
\label{Ssec: optical}
Five of the \madcows\ clusters have follow-up $r$- and $z$-band
imaging with the Gemini Multi-Object Spectrograph
\citep[GMOS,][]{GMOS} on Gemini-North with five 180 s exposures in the
$r$-band and twelve 80 s exposures in the $z$-band from programs
GN-2013A-Q-44 and GN-2013B-Q-8 (PI Brodwin). The data were taken
between 2013 February and 2015 July.


\subsection{New Spectroscopic Redshifts}
\label{Ssec: speczs}
Five of the \madcows\ clusters presented here have previously
unreported spectroscopic redshifts. We obtained spectroscopic
observations of these five clusters from 2015 through 2017, primarily
with the Low Resolution Imaging Spectrometer \citep[LRIS,][]{LRIS} at
the W.~M.~Keck Observatory, the details of which are given in Table
\ref{Table: all speczs}. The mask used for each cluster was designed
from the \spitzer\ imaging and focused on the IRAC sequence members
identified in a \cho$-$\cht\ color-magnitude diagram.

\begin{figure}[bthp]
\center{\includegraphics[width=8.5cm]{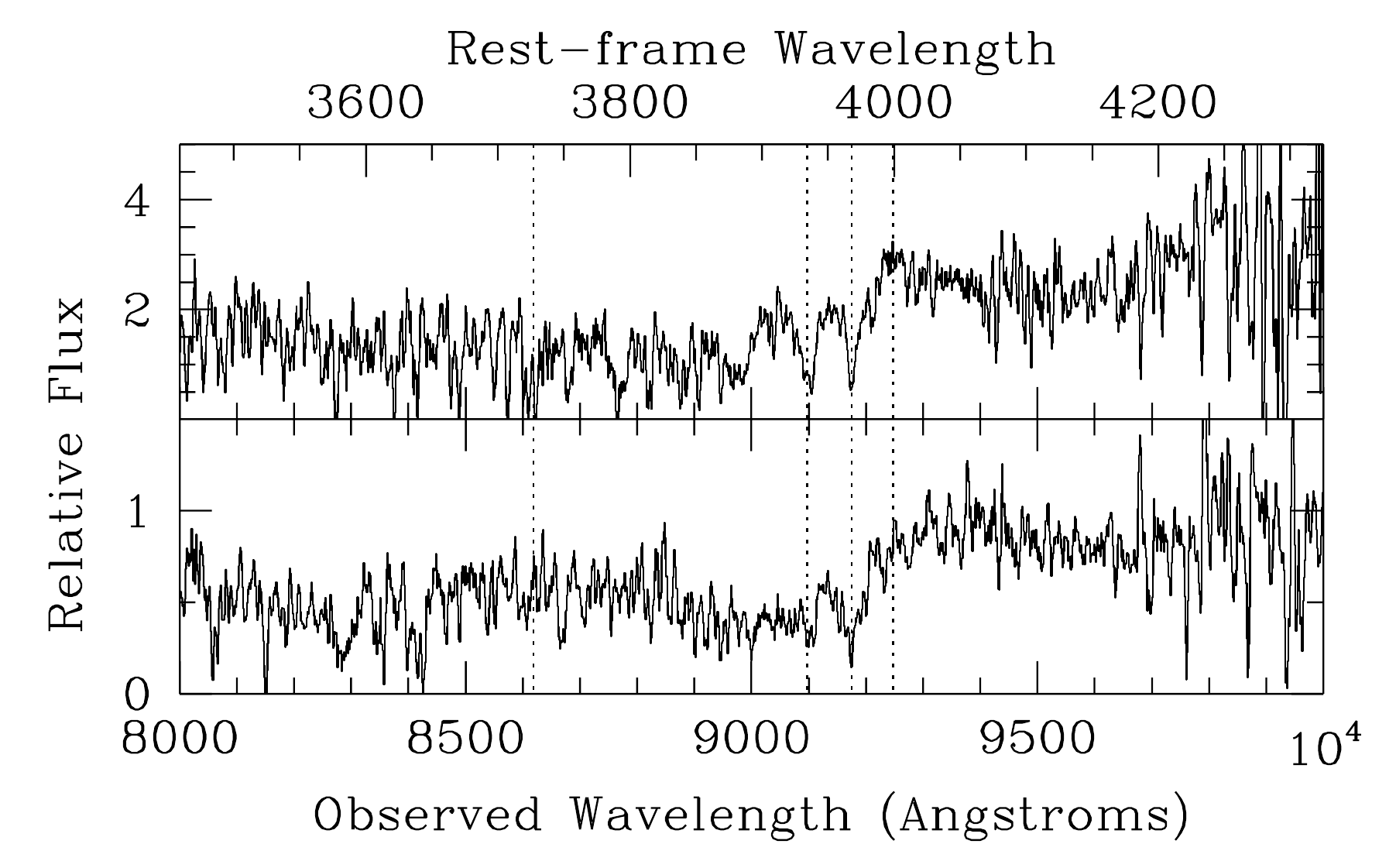}}
\caption{Spectra of two of the six confirmed members of MOO\
J1521+0452 at $z=1.31$. The vertical lines show , left-to-right
the locations of [O II] $\lambda 3727$, Ca II K \& H lines and D4000.} 
\label{Fig: MOO1521 spectra}
\end{figure}

One of the clusters with new spectroscopic redshifts reported here,
MOO\ J1521+0452, is the highest-redshift \madcows\ cluster with
spectroscopy, and with $M_{500} =
(3.59^{+1.02}_{-0.92})\teneft~\msol$, it is the \rank-most massive
cluster to be found at $z\ge1.3$ by any method. The spectroscopy
confirmed six cluster members and established $z=1.312$ as the cluster
redshift. Representative spectra of two of the confirmed members are
shown in Figure \ref{Fig: MOO1521 spectra}.

Four cluster members were confirmed for MOO\ J0037$+$3306,
establishing the cluster redshift of $z=1.133$. MOO\ J0105$+$1323 and
MOO\ J0123$+$2545 each had five identified members, placing their
redshifts at $z=1.143$ and $z=1.215$, respectively.

In addition to the newly reported clusters above, we also present a
new spectroscopic redshift for MOO\ J1014+0038, previously reported at
a photometric redshift of $z_\mathrm{phot}=1.27\pm0.08$
\citep{Brodwin+15}. In addition to LRIS spectroscopy, we also observed
this cluster with the Multi-Object Spectrometer For Infra-Red
Exploration \citep[MOSFIRE,][]{MOSFIRE10,MOSFIRE12} at Keck on 2016
February 01. These new spectra identified seven members and
established the redshift for MOO\ J1014+0038 as $z=1.231$. Spectra
for two of these members are shown in Figure \ref{Fig: MOO1014
spectra}.


\begin{figure}[bthp]
\center{\includegraphics[width=8.5cm]{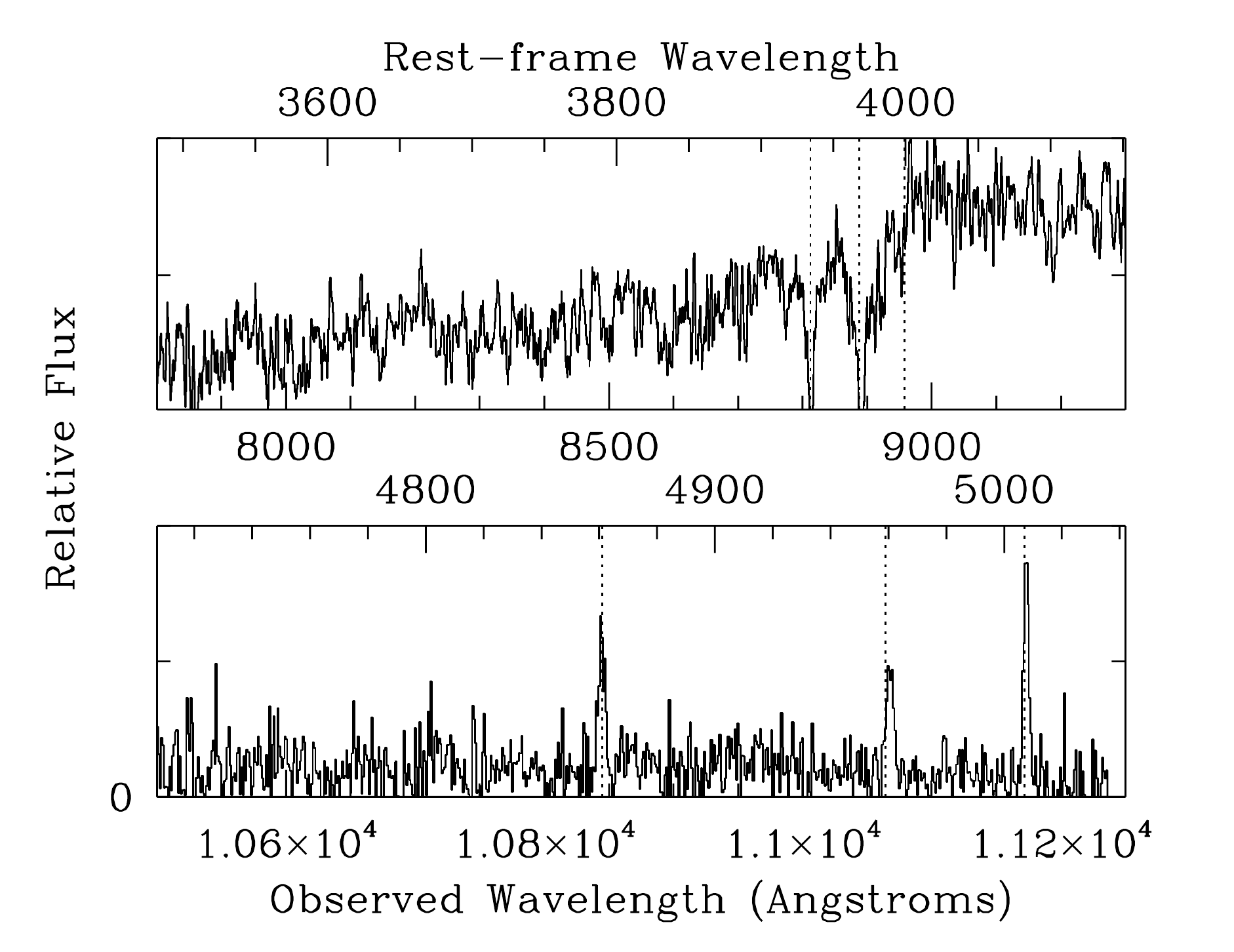}}
\caption{Spectra of confirmed MOO\ J1014+0038 cluster members from
LRIS (top) and MOSFIRE (bottom), establishing a cluster redshift of
$z=1.231$. Left-to-right, the vertical lines of the top spectrum
show the Ca II K\&H lines and the 4000 \AA\ break and the vertical
lines of the lower plot show the H$\beta$ and [O III] emission
features.}
\label{Fig: MOO1014 spectra}
\end{figure}


\begin{deluxetable*}{llccccc}
  \tabletypesize{\normalsize} \tablecaption{Spectroscopic Cluster
    Members\label{Table: all speczs}} \tablewidth{0pt} \tablehead{
    \colhead{RA} & \colhead{Dec.} & \colhead{Instrument} & \colhead{UT
    Date} & \colhead{$z$} &
    \colhead{Quality\tablenotemark{a}} & \colhead{Features} } \startdata
  \cutinhead{MOO\ J0037$+$3306 $z=1.133$}
  00:37:45.77 & $+$33:07:50.9 & LRIS & 2016 August 05 & 1.131 & A & D4000 \\
  00:37:46.18 & $+$33:07:28.2 & LRIS & 2016 August 05 & 1.123 & A & Ca HK \\
  00:37:48.82 & $+$33:07:08.4 & LRIS & 2016 August 05 & 1.15 & B & D4000 \\
  00:37:47.03 & $+$33:06:45.7 & LRIS & 2016 August 05 & 1.13 & B & D4000 \\
  \cutinhead{MOO\ J0105$+$1323 $z=1.143$}
  01:05:26.64 & $+$13:23:36.9 & LRIS & 2015 December 04 & 1.13 & B & D4000 \\
  01:05:26.20 & $+$13:23:53.7 & LRIS & 2015 December 04 & 1.14 & B & D4000 \\
  01:05:29.95 & $+$13:23:54.6 & LRIS & 2015 December 04 & 1.15 & A & Ca HK,D4000 \\
  01:05:35.27 & $+$13:23:10.4 & LRIS & 2015 December 04 & 1.144 & B & [O II]$\lambda$3727,D4000 \\
  \cutinhead{MOO\ J0123$+$2545 $z=1.215$}
  01:23:50.95 & $+$25:45:47.19 & LRIS & 2017 July 20 & 1.20 & B & D4000 \\
  01:23:57.16 & $+$25:44:16.67 & LRIS & 2017 July 20 & 1.22 & B & D4000 \\
  01:23:47.37 & $+$25:46:50.65 & LRIS & 2017 July 20 & 1.2214 & A & [O II]$\lambda$3727 \\
  01:23:41.53 & $+$25:47:32.78 & LRIS & 2017 July 20 & 1.2196 & B & [O II]$\lambda$3727 \\
  \cutinhead{MOO\ J1014$+$0038 $z=1.231$}
  10:14:07.31 & $+$00:38:27.1 & LRIS & 2015 February 21 & 1.231 & B & Ca HK \\
  10:14:10.51 & $+$00:37:56.2 & LRIS & 2015 February 21 & 1.23 & B & D4000 \\
  10:14:08.11 & $+$00:37:36.6 & LRIS & 2015 February 21 & 1.239 & A & Ca HK \\
  10:14:00.32 & $+$00:36:43.7 & LRIS & 2015 February 21 & 1.22 & B & [O II]$\lambda$3727 \\
  10:14:08.13 & $+$00:38:21.3 & LRIS & 2015 December 06 & 1.23 & B & Ca HK,D4000 \\
  10:14:12.80 & $+$00:38:12.2 & MOSFIRE & 2016 February 01 & 1.2318 & A & H$\beta$,[O III]$\lambda$4959,5007 \\
  10:14:09.71 & $+$00:41:11.1 & LRIS & 2016 March 06 & 1.226 & B & [O II]$\lambda$3727 \\
  \cutinhead{MOO\ J1521$+$0452 $z=1.312$}
  15:21:13.66 & $+$04:53:28.0 & LRIS & 2016 July 05 & 1.308 & B & Ca HK \\
  15:21:12.10 & $+$04:51:16.9 & LRIS & 2016 July 05 & 1.317 & B & Ca HK \\
  15:21:06.79 & $+$04:52:09.1 & LRIS & 2016 July 05 & 1.312 & B & Ca HK,D4000 \\
  15:21:04.90 & $+$04:51:59.8 & LRIS & 2016 July 05 & 1.302 & B & Ca HK,D4000 \\
  15:21:04.15 & $+$04:52:12.4 & LRIS & 2016 July 05 & 1.32 & B & Ca HK,D4000 \\
  15:20:59.35 & $+$04:51:40.7 & LRIS & 2016 July 05 & 1.314 & A & Ca HK,D4000 \\
  \cutinhead{Foreground/Background Objects}
  00:37:51.56 & $+$33:10:07.0 & LRIS & 2016 August 05 & 1.453 & A & [O II]$\lambda$3727,D4000 \\
  01:05:22.72 & $+$13:23:55.2 & LRIS & 2015 December 04 & 0.229 & A & [O II]$\lambda$3727 \\ 
  01:05:35.14 & $+$13:23:36.9 & LRIS & 2015 December 04 & 0.248 & A & [O II]$\lambda$3727,H$\alpha$,H$\beta$ \\
  01:23:48.16 & $+$25:46:01.2 & LRIS & 2017 July 20 & 0.2120 & A & H$\alpha$,H$\beta$ \\
  01:23:42.28 & $+$25:46:31.4 & LRIS & 2017 July 20 & 0.4659 & A & H$\alpha$,[O III]$\lambda$4959,5007 \\
  01:23:42.32 & $+$25:47:17.5 & LRIS & 2017 July 20 & 0.4364 & A & H$\alpha$ \\
  01:23:56.71 & $+$25:46:31.7 & LRIS & 2017 July 20 & 1.4781 & A & [O II]$\lambda$3727,Ca HK \\
  10:14:11.57 & $+$00:38:39.3 & LRIS & 2015 February 21 & 1.158 & A & Ca HK,D4000 \\
  10:14:02.48 & $+$00:34:53.0 & LRIS & 2015 February 21 & 0.326 & A & H$\alpha$ \\
  10:14:13.36 & $+$00:39:57.8 & LRIS & 2015 December 06 & 0.966 & A & Ca HK,D4000,G \\
  10:14:04.15 & $+$00:41:03.5 & LRIS & 2015 December 06 & 0.981 & A & [O II]$\lambda$3727, Ca HK,D4000 \\
  10:14:00.76 & $+$00:40:23.2 & LRIS & 2015 December 06 & 0.3283& A & H$\alpha$,[N II],Na D \\
  15:21:08.78 & $+$04:52:59.5 & LRIS & 2016 July 05 & 0.514 & A & H$\alpha$,[N II] \\
  15:20:52.34 & $+$04:51:32.0 & LRIS & 2016 July 05 & 0.489 & A & H$\alpha$,[N II]
  \enddata
  \tablenotetext{a}{Qualities `A' and `B' denote redshifts of high
    and reasonable certainty, respectively \citep{Stanford+14}. }
\end{deluxetable*}

\section{Analysis}
\label{Sec: analysis}
\subsection{Total Cluster Mass}
\label{Ssec: mass}
Details of the CARMA observations are given in Table \ref{Table: Carma
observations}. The data, including those for clusters previously
reported in \citet{Brodwin+15} and \citet{Gonzalez+15}, were
re-reduced using a new MATLAB pipeline designed specifically for CARMA
data. Mars was used as the flux calibrator for each cluster with the
\citet{Rudy+87} flux model and observations of a bright monochromatic
quasar were interleaved with the cluster observations for gain
calibration. The cluster Comptonization ($Y_\mathrm{SZ}$) was
calculated by using a Monte Carlo Markov Chain to fit an
\citet{Arnaud+10} pressure profile and point source models (where
indicated by the long baseline data) to the CARMA data in {\it uv}
space. The significance of the detection was calculated by comparing
$\chi^2$ for the fit to the Arnaud model and point source(s) to
$\chi^2$ for a fit to just the point source(s) with no cluster
model. \mfh\ and \rfh\ were calculated from $Y_{SZ}$ by forcing
consistency with the scaling relation from \citet{Andersson+11}. The
resulting masses, radii and $Y_{SZ}$ values are shown in Table
\ref{Table: Carma observations}. Updated masses and radii, based on
the new pipeline, are reported for the clusters reported in
\citet{Brodwin+15} and \citet{Gonzalez+15}. These are all consistent
within one sigma with the originally reported quantities. The
total masses for the SPT-SZ sample are from the \citet{Bleem+15}
catalog.


\subsection{Catalogs}
\label{Ssec: catalogues}
For each cluster we ran SExtractor \citep{SExtractor} in dual-image
mode on the \cho\ and \cht\ images, selecting on the \cho\ image. We
used the IRAC coverage maps as weights and SExtractor parameters
similar to those in \citet{Lacy+05}. These parameters are optimized
for IRAC, but we changed DEBLEND\_NTHRESH to 64 and DEBLEND\_MINCONT
to 0.00005 to better deblend sources in the cluster cores. Magnitudes
were measured in $4\arcsec$ diameter apertures and corrected to
$24\arcsec$ diameter apertures using the corrections from
\citet{Ashby+09}. Catalogs for the optical images were produced with
the same SExtractor parameters, but with MAG\_AUTO magnitudes instead
of corrected aperture magnitudes. The optical and infrared catalogs
were then matched using the IRAC astrometry to produce combined
catalogs for each cluster. All of the catalogs have IRAC \cho\
and \cht\ fluxes that are $\ga70\%$ complete down to magnitudes of
21.0 and 22.5, respectively. The clusters with optical data have
additional $r$- and $z$-band data similarly complete to depths of 25.5
and 24.5 magnitudes.


\subsection{Cluster Membership}
\label{Ssec: membership}
\begin{figure}[bthp]
\center{\includegraphics[width=8.5cm]{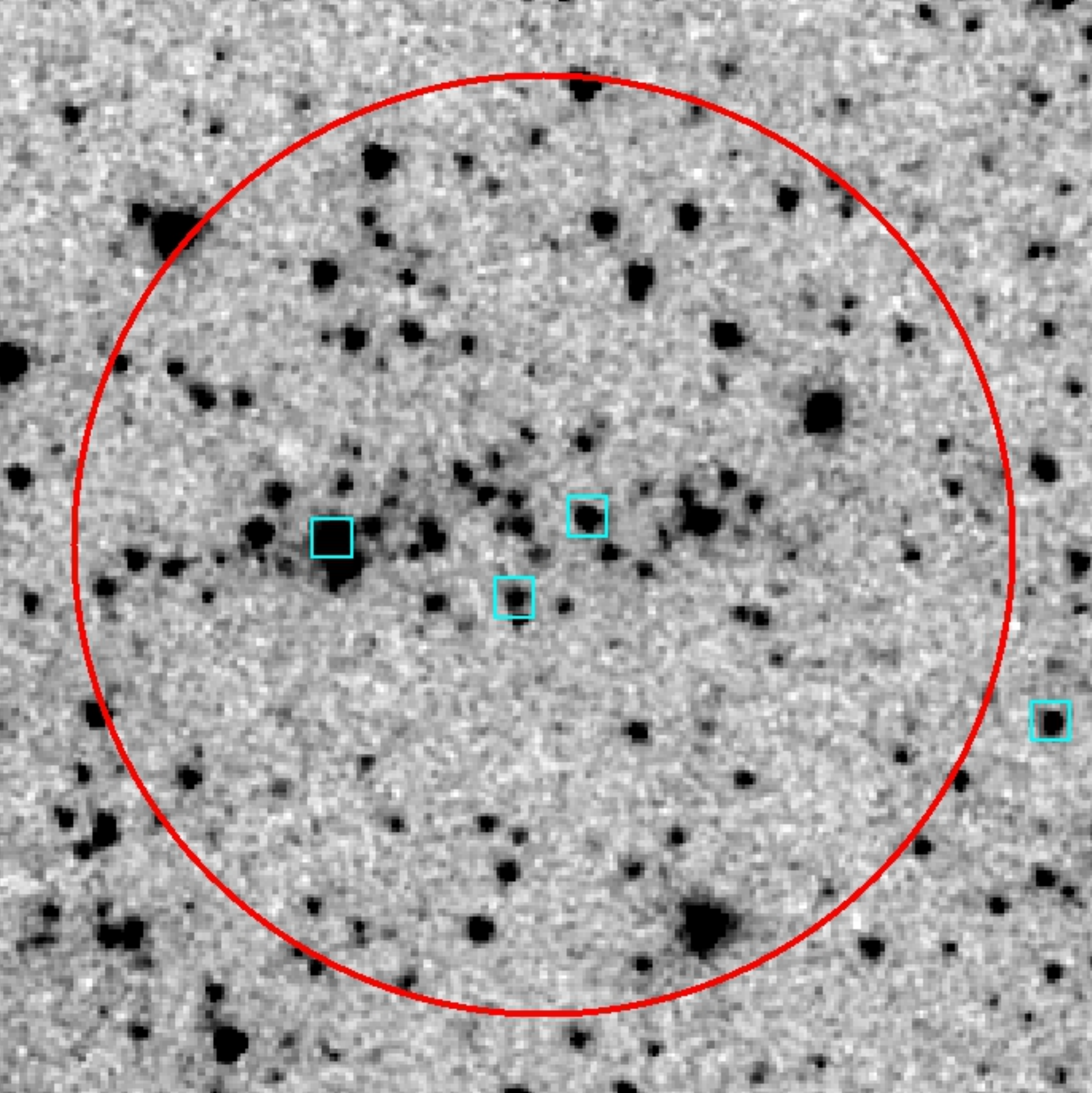}}
\caption{A $170\arcsec \times 170\arcsec$ \spitzer/IRAC \cho\ image of
MOO J1521+0452 showing an $80\arcsec$ radius circle corresponding to
the cluster \rfh\ of 0.67 Mpc at $z=1.31$. Only galaxies inside the
red circle were included in the analysis. Cluster members with
spectroscopic redshifts are marked with cyan squares.}
\label{fig_MOO_r500} 
\end{figure}

Because our cluster masses are measured at an overdensity of $\Delta =
500$, we only consider galaxies projected within \rfh\ (as determined
from the SZ data) from the centroid of the SZ decrement in our
measurement of stellar masses and fractions (\eg\ Figure
\ref{fig_MOO_r500}). To ensure our choice of center does not
significantly impact our results, we also ran our analysis using the
centroid of the galaxy distribution and using the BCG as the
center. We find no appreciable differences in our results. Within
\rfh, we also reject objects that likely lie in the foreground by not
including any source with an apparent magnitude brighter than $\mstar
- 2$ at the redshift of our cluster. The effects of this choice
of cutoff are discussed in \textsection{\ref{Ssec: systematics}}. The
characteristic magnitude was calculated using the same model as was
used for our K-corrections (described in \textsection\ref{Ssec:
stellar mass}). To limit the effect of incompleteness at the
faint-end, we reject objects more than one magnitude fainter than
$\mstar$.

\begin{figure}[bthp]
\center{\includegraphics[width=8.5cm]{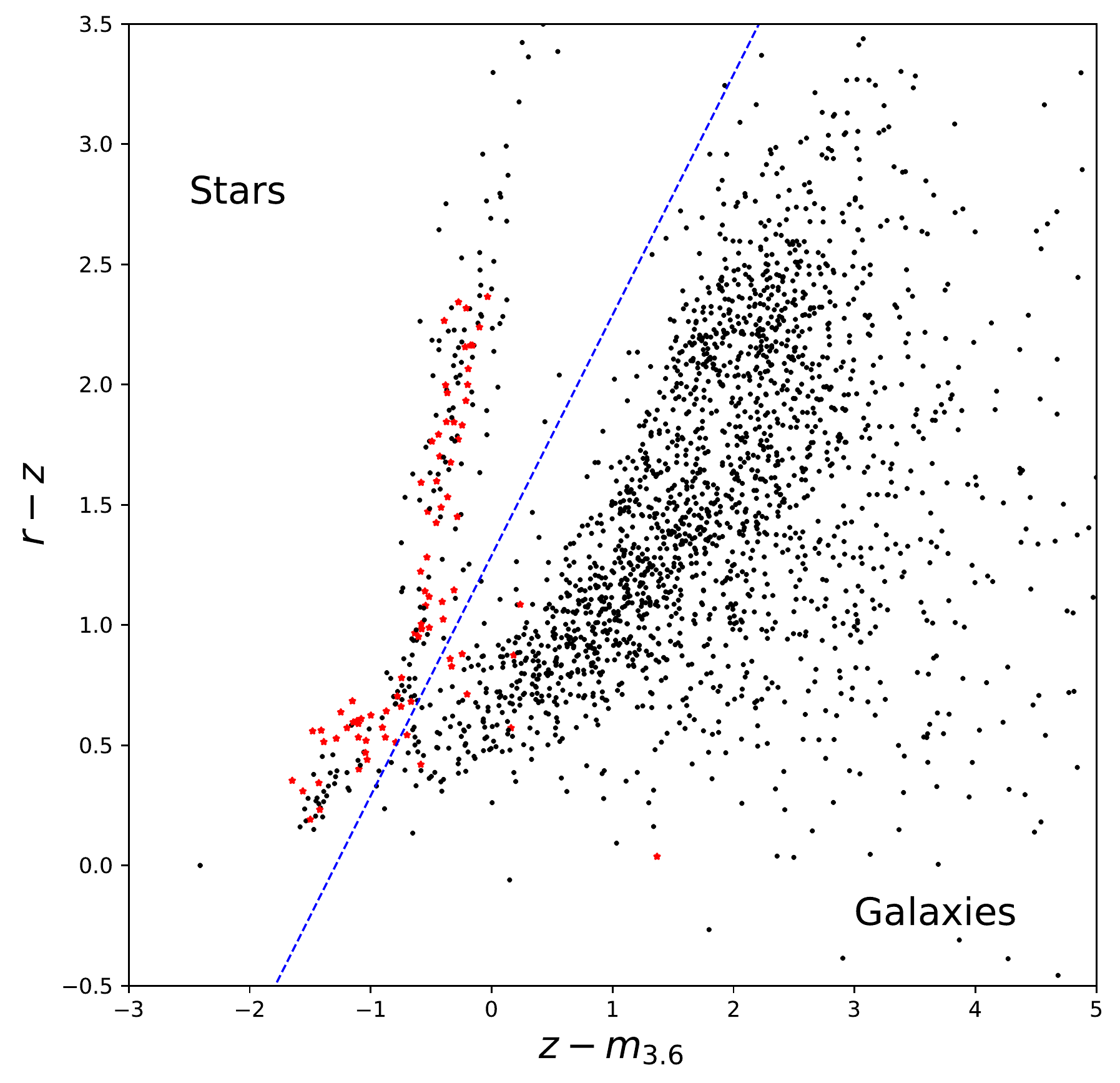}}
\caption{Combined color-color plot of all the \madcows\ clusters for
which there are GMOS data, showing $r - z$ color plotted against $z
-m_{3.6}$. Objects above the blue dashed line have colors consistent
with being stars. Objects with $\ge3~\sigma$ parallax in the {\it
Gaia} DR2 catalog are plotted as red stars.}
\label{Fig: stellar_rejection} 
\end{figure}

We used the available optical data for \nopt\ of the \madcows\
clusters to identify stars in color-color space. Following
\citet{Eisenhardt+04}, we plot $r - z$ versus $z-$\cho\ colors for
each of our possible cluster members. To the limit where our optical
data are complete for all clusters, we characterize as stars objects
falling above the line shown in Figure \ref{Fig: stellar_rejection}
that separates objects with the colors of stars from objects that are
likely galaxies. Only objects bright enough to be clearly detected in
even the shallowest of our optical images are so characterized to
ensure a consistent cut across all clusters. We also match
our catalogs to objects in the {\it Gaia} DR2 catalog
\citep{Gaia16,GaiaDR2} with greater than $3~\sigma$ parallax, to
confirm that objects known to be stars are the objects being rejected
by this approach. We cannot do this for the SPT clusters due to a
lack of comparable optical data.

Although the bulk of galaxies within \rfh\ are cluster members, there
is still a line-of-sight interloper contribution that must be
subtracted. To account for this, we determine the expected
contribution to the total flux density from field galaxies within the
projected \rfh\ area and subtract it off the flux density calculated
from our cluster. We use the \spitzer\ Deep Wide-Field Survey
\citep[SDWFS,][]{Ashby+09} to do this, applying the same brightness
cuts to reject non-cluster members as we apply to our cluster
catalogs. For the clusters with optical data allowing the rejection of
stars, we use optical photometry from the NOAO Deep Wide-Field Survey
\citep[NDWFS,][]{NDWFS} to make the analogous stellar rejection in our
background. For each cluster, we treat all remaining objects in the
SDWFS catalog as though they were at the redshift of that cluster and
calculate how much spurious luminosity they would add. We use the
SDWFS field to determine our background because the IRAC imaging is
deeper than that of our clusters and because SDWFS is large enough to
smooth out small-scale variations in the background level. This
background selection methodology does produce an appreciable
systematic uncertainty in our results, as discussed in
\textsection\ref{Ssec: systematics}.

\subsection{Completeness}
\label{Ssec: completeness}
To correct for incompleteness in our IRAC catalogs, we ran
completeness simulations over the range of magnitudes at which we were
looking using IRAF's {\it mkobjects} task in the {\it noao}/{\it
artdata} package. For each cluster we added ten random point sources
in each half magnitude bin to the IRAC \cho\ image, and ran SExtractor
to see how many were recovered. We repeated this process 1,000 times
in each magnitude bin. This was done for both the \madcows\ and SPT
clusters and we performed a similar analysis on the SDWFS \cho\ image
and on the optical images of the clusters. The average completeness
curve for \madcows\ and SPT are shown in Figure \ref{Fig:
Completeness}. At $\mstar+1$, the faint-end limit of our analysis, the
catalogs of both surveys are approximately 70\% complete, depending
slightly on cluster redshift. Because our clusters have slightly
different $\mstar$ (depending on redshift), our faint-end cutoff
varies slightly, as shown in the figure.

\begin{figure}[bthp]
\center{\includegraphics[width=8.5cm]{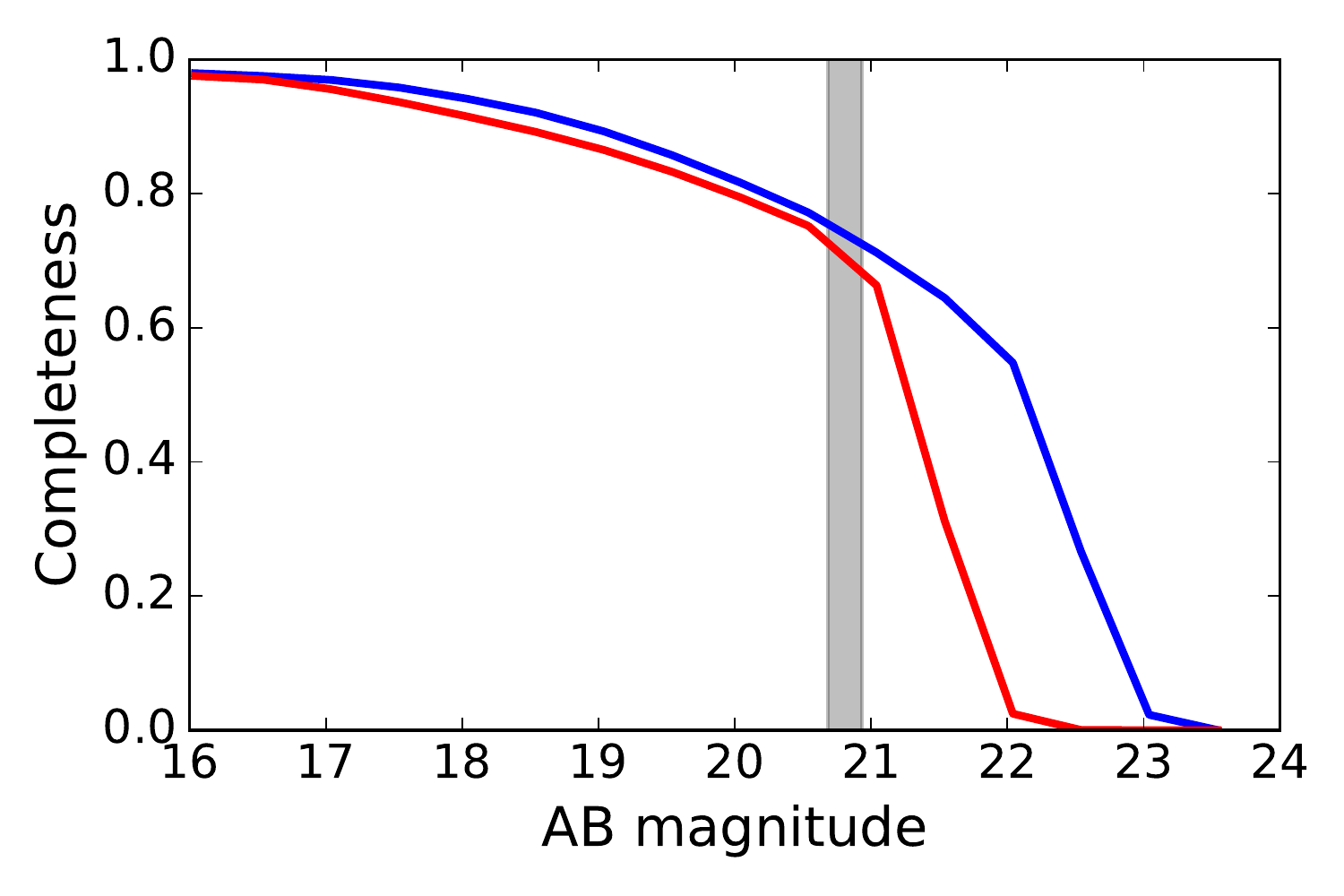}}
\caption{Average completeness curves for the \madcows\ (red) and SPT
(blue) IRAC \cho\ images in AB magnitudes. The shaded region
represents the range of maximum depths to which our analysis extends.}
\label{Fig: Completeness}
\end{figure}


\section{Results} 
\label{Sec: results}
\subsection{Stellar Mass}
\label{Ssec: stellar mass}
We calculate the stellar mass of the galaxies selected as possible
cluster members using their rest-frame $H$-band luminosity. The rest
frame $H$-band is centered at the peak of the emission from the old,
red stars that dominate the stellar mass of the galaxy. It is
therefore a relatively low-scatter proxy for total stellar mass
\citep[\eg][]{Hainline+11} with a relatively small dependence on the
overall SED. At $z\sim1$ this is easily probed by the IRAC \cho\ band.
To determine the K-correction from observed IRAC \cho\ to rest-frame
$H$-band, we use EZGal \citep{EZGal}. We construct a synthetic galaxy
SED with a \citet{BC03} $1~\mathrm{Gyr}$ tau model, formation redshift
$z_\mathrm{f}=3$, solar metallicity and a \citet{Chabrier03} IMF. From
this SED we derive a K-correction to the absolute magnitude in the
$H$-band, from which we calculate $L_\mathrm{H}$. We statistically
correct our luminosities for incompleteness using the simulations
described above. We use the same EZGal model to determine the stellar
mass-to-light ratio in the $H$-band at the cluster redshift. This
$M/L$ ratio is different for each cluster, depending on the redshift,
but is close to 0.34 on average. We apply the stellar $M/L$ ratio to
the sum of the luminosities of all the objects along the line of sight
minus the background contribution estimated from SDWFS to get our
final cluster stellar mass. Both the \madcows\ and SPT clusters were
analyzed in the same way and to the same depth to allow for direct
comparison of the two samples.


\subsection{Estimating Stellar Corrections with Luminosity Functions}
\label{Ssec: LF}
Before calculating total stellar mass fractions, we need to account
for foreground stars along the line-of-sight to our clusters. We do
this by combining the optical stellar identification discussed above
with cluster luminosity functions to estimate and correct the total
impact from stars on our clusters that lack optical data.

\begin{figure}[bthp]
\center{\includegraphics[width=8.5cm]{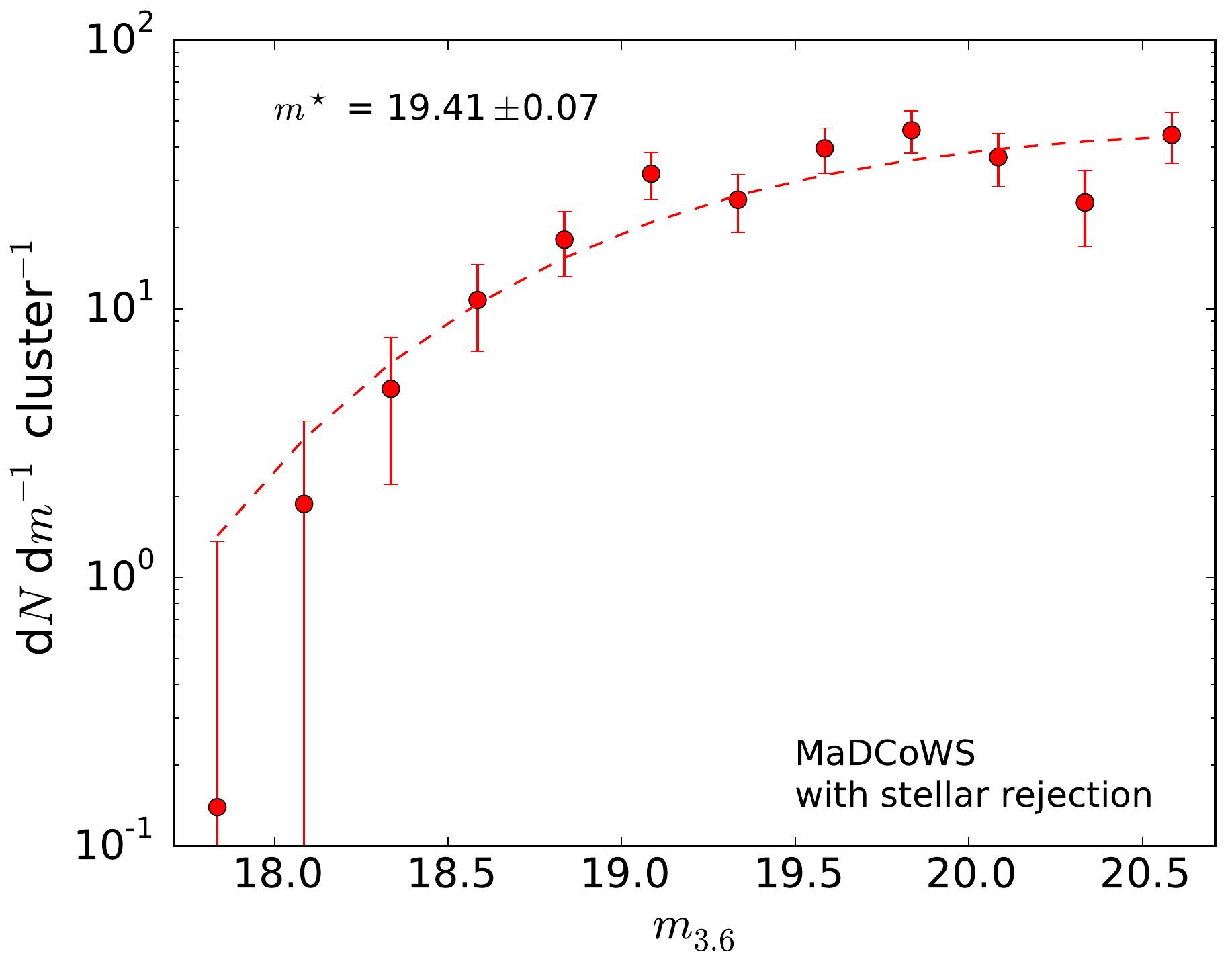}}
\caption{The IRAC \cho\ luminosity function for the \nopt\ \madcows\
clusters with optical data for stellar rejection. The solid circles
are background subtracted number per magnitude in each bin and the
error bars are from Poisson noise. The dashed line is a best-fit
Schechter function with a fixed $\alpha=-0.8$ and the best-fit value
of $\mstar$ is shown.}
\label{Fig: madcows LF}
\end{figure}

The mean IRAC \cho\ luminosity function (LF) for the \nopt\ \madcows\
clusters with optical data for stellar rejection is shown in Figure
\ref{Fig: madcows LF}.  To make this LF we applied the membership cuts
from \textsection{\ref{Ssec: membership}}, including stellar rejection
from the optical data, to each cluster and evolution-corrected the
members to $z=1$. The galaxies from all the clusters were then
binned in quarter-magnitude wide bins and the appropriate completeness
and statistical background corrections were applied. The uncertainties
on the values are Poisson errors.

We fit to the data a parameterized Schechter function of the
form $$\Phi(m) = 0.4\ln(10)\Phi^*
10^{-0.4(m-\mstar)(\alpha+1)}\mathrm{exp}(-10^{-0.4(m-\mstar)}) $$
\citep{Schechter76} and we fix $\alpha=-0.8$ as our data are not deep
enough to constrain the faint-end slope. This choice is consistent
with \citet{Mancone+10} and is a reasonable value for our data. The
best-fit value is $\mstar = \fitms\pm0.07$ and we take this LF as
representative of $z=1$ clusters independent of selection. The error
on the $\mstar$ fit is calculated from the range of $\chi^2$, and is
the same as the error calculated from bootstrap resampling. This
value for $\mstar$ is slightly lower than, but close to that of
\citet{Muzzin+08} who found $\mstar_{3.6} = 20.11\pm0.64$ (in AB
magnitudes) for IRAC \cho\ at $z=1.01$ and \citet{Mancone+10} who
found $\mstar_{3.6} = 19.71\pm0.06$ at $z=0.97$. It is also consistent
with the value of $\mstar_{3.6} = 19.62^{+0.25}_{-0.20}$ found for
infrared-selected clusters in a higher redhift bin ($z=1.45$) by
\citet{Wylezalek+14}.

\begin{figure}[bthp]
\center{\includegraphics[width=8.5cm]{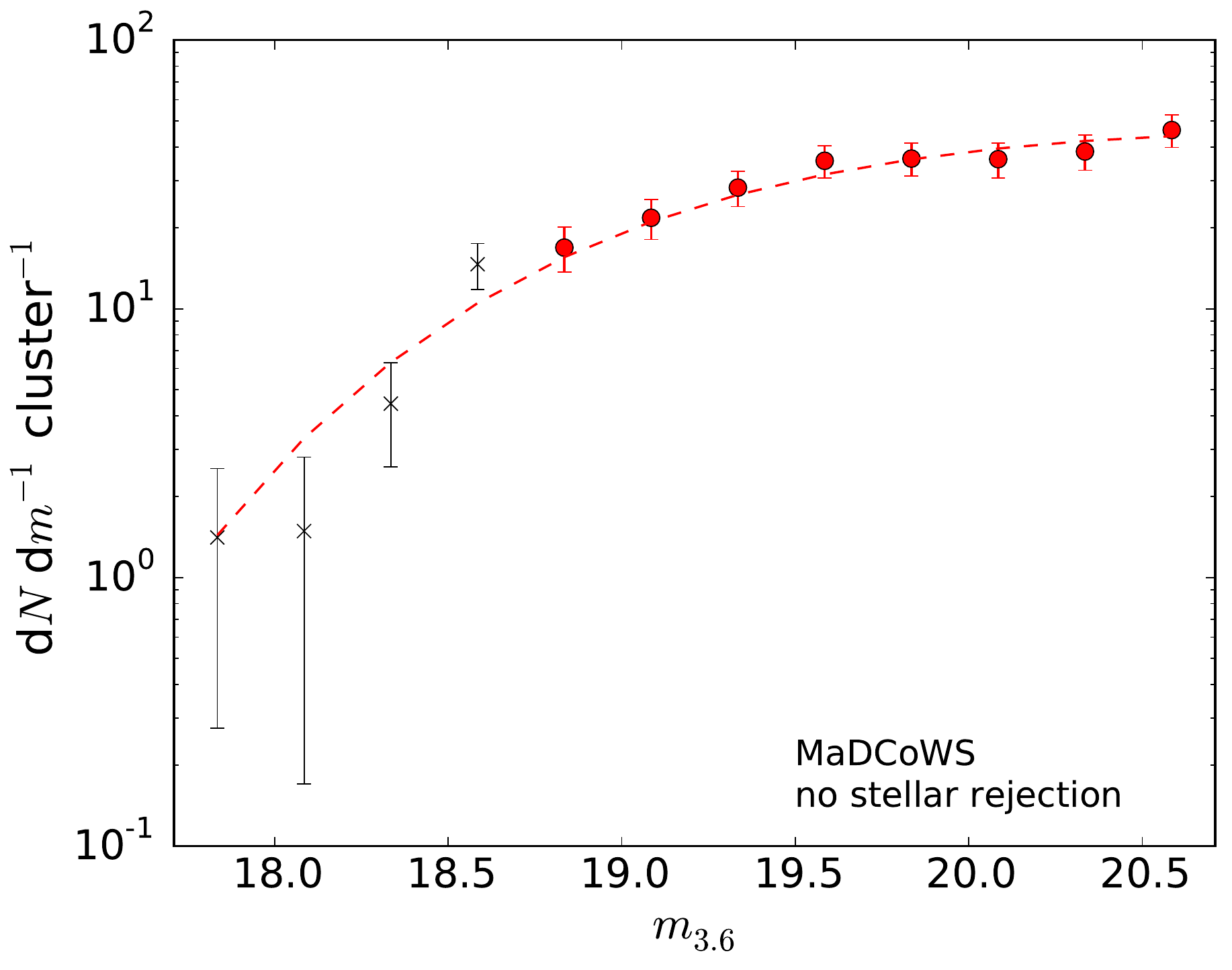}}
\caption{The mean IRAC \cho\ luminosity function for the full sample
  of twelve \madcows\ clusters with no optical rejection of
  stars. All of the points are background-subtracted number per
 magnitude in each bin and the error bars are from Poisson noise. The
 black crosses on the bright end are points with potential stellar
 contamination that we did not include when fitting the Schechter function,
 which is represented by the dashed red line. For the Schechter function, we
 fixed $\alpha=-0.8$ and $\mstar = \fitms$ to match the luminosity
 function derived using optical stellar rejection.} 
\label{Fig: all madcows LF}
\end{figure}

\begin{figure}[bthp]
\center{\includegraphics[width=8.5cm]{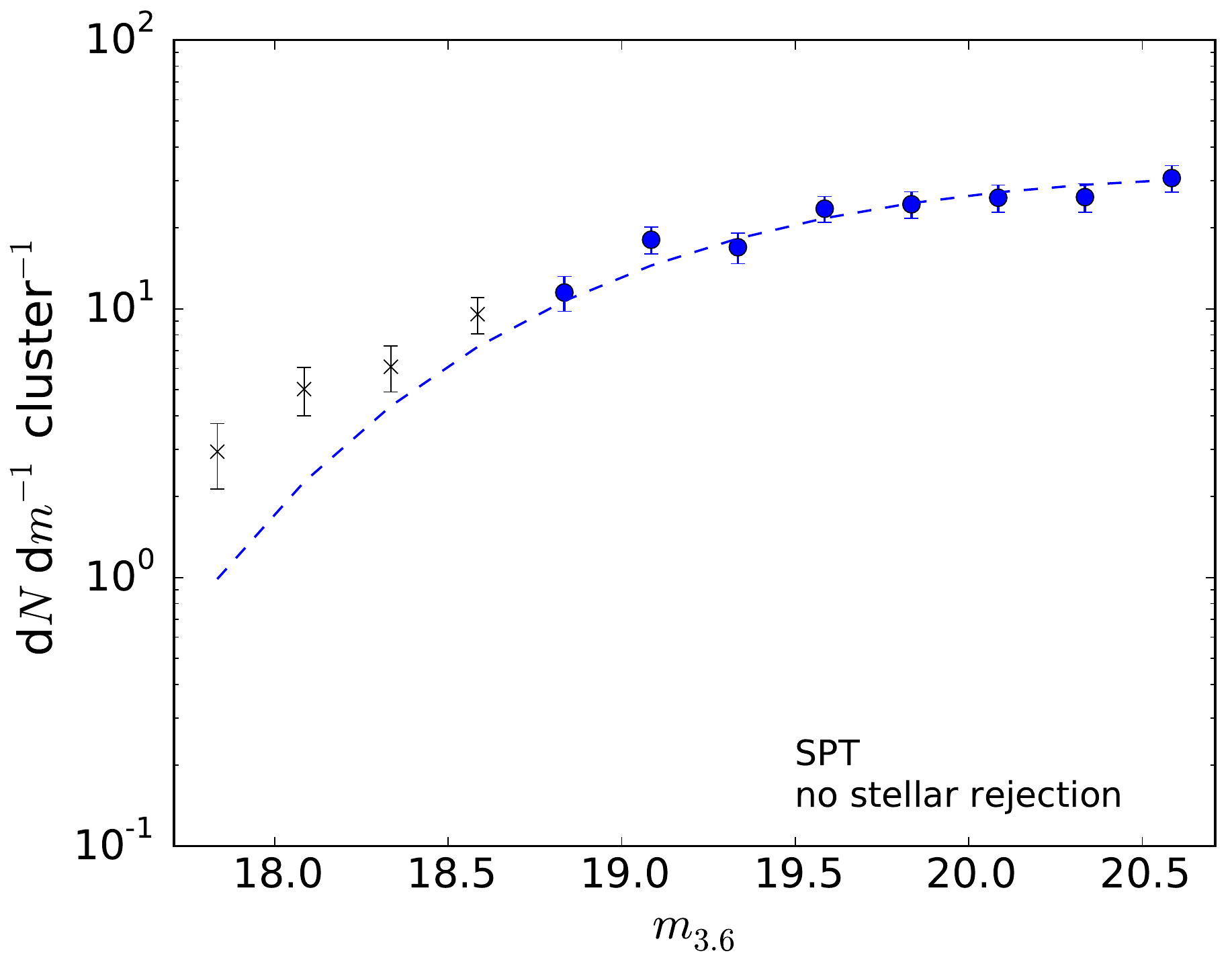}}
\caption{The mean IRAC \cho\ luminosity function for the \nSPT\
comparison SPT clusters in this work. All of the points are
background-subtracted number per magnitude in each bin and the error
bars are from Poisson noise. The black crosses on the bright end are
points with likely stellar contamination that we did not include when
fitting the Schechter function, which is represented by the dashed
blue line. For the Schechter function, we fixed $\alpha=-0.8$ and
$\mstar = \fitms$ to match the \madcows\ LF. Note the stellar
contamination at the bright end.}
\label{Fig: SPT LF}
\end{figure}

We used a similar approach to make luminosity functions for the full
sample of twelve \madcows\ clusters and for the SPT clusters, shown in
Figures \ref{Fig: all madcows LF} and \ref{Fig: SPT LF},
respectively. The stellar contamination is more extensive for the SPT
clusters because the sample extends to a lower galactic latitude,
where there is more line-of-sight contamination, than does the
\madcows\ sample. We do not have adequate optical data for all of
these clusters and thus do not attempt stellar corrections on a
per-cluster basis. Rather, we construct a statistical stellar
correction as follows. We fit the $z=1$ Schechter function determined
above, allowing only $\Phi^*$ to vary (\ie\ with fixed $\alpha = 0.8$
and $\mstar = \fitms$ as for the clusters without stellar
contamination), to the points at the faint end of \madcows\ and SPT
LFs that show no evidence of stellar contamination (as determined by
the SPT LF). These are the points plotted with filled circles in
Figures \ref{Fig: all madcows LF} and \ref{Fig: SPT LF}; the unfit
portion of the LFs, where there appears to be significant stellar
contamination in the SPT LF, is plotted with black crosses. The ratios
between the areas under these `no-stars' fits for each sample, over
the full magnitude range in this work, to the area under their
respective observed LFs is the statistical stellar correction factor
for that sample. We multiply the measured luminosity of each cluster
by the correction factor of the sample to get the true luminosity for
that cluster absent stellar contamination.

\subsection{Stellar Mass Fraction}
\label{Ssec: fstar}
\begin{figure*}[bthp]
\center{\includegraphics[width=18cm]{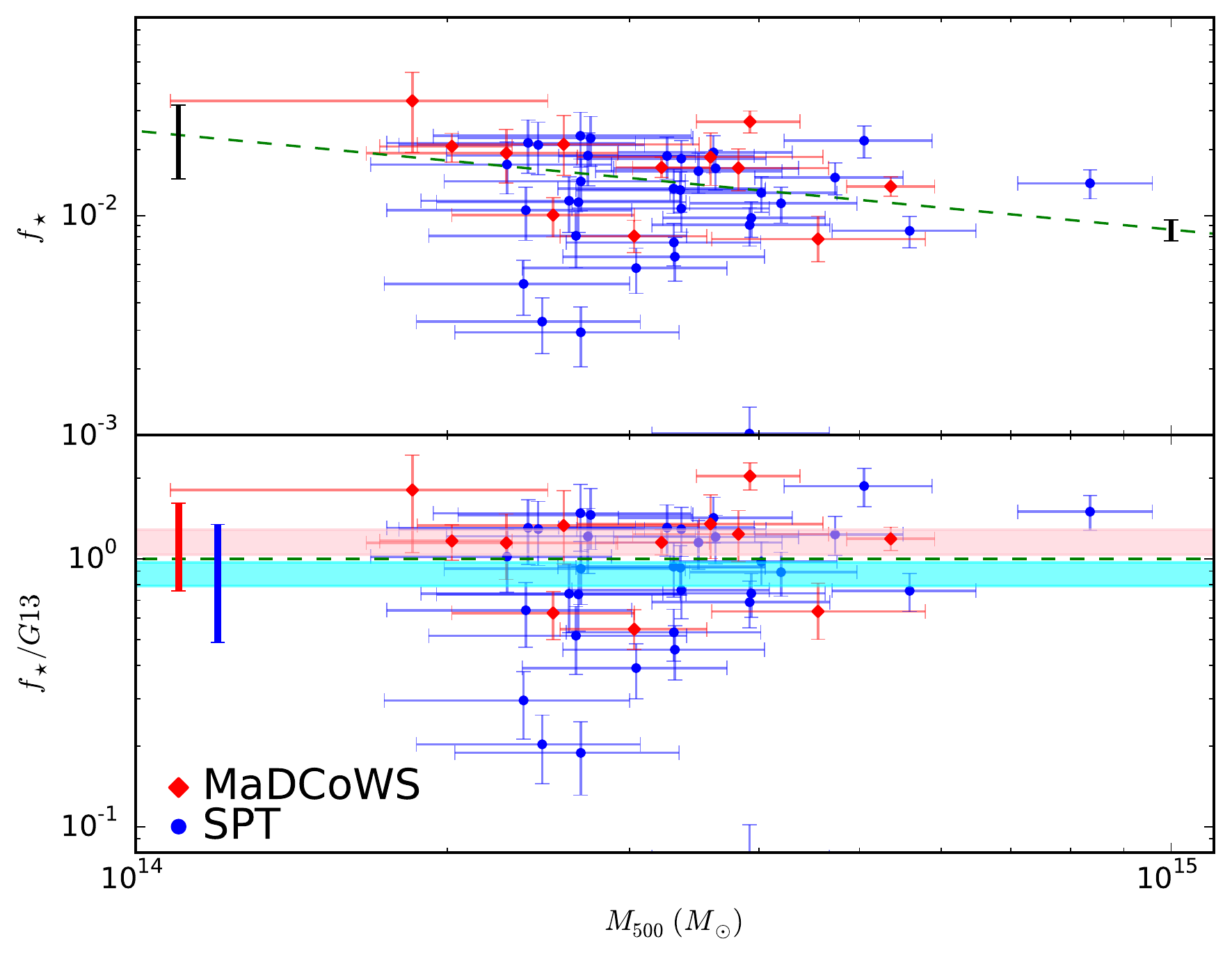}}
\caption{{\it Upper}: Stellar mass fraction versus total mass for the
\madcows\ (red) and SPT (blue) clusters. The size of the systematic
error in \fstar, which varies with \mfh, is represented by the black
error bars on either end of the figure. The green dashed line is the
low-redshift relation from G13.
{\it Lower}: The stellar mass fractions of each cluster normalized by
the G13 relation versus total mass. The
error about the mean normalized \fstar\ for both samples is calculated
from bootstrap resampling and for \madcows\ (SPT) is
plotted in pink (cyan) across the figure. The scatter in
normalized \fstar\ is shown by the thick, vertical red (blue) error bars.}
\label{Fig: fstarvm500}
\end{figure*}

To calculate \fstar, we divide the stellar mass of the cluster by
the total mass calculated from the SZ decrement described above. The
stellar mass that we use is calculated by summing the completeness-
and K-corrected $H$-band luminosity of every object projected within
\rfh\ of the cluster SZ centroid and subtracting the average
background calculated from SDWFS. We then multiply this luminosity by
the M/L ratio from our EZGal model for the cluster redshift and the
average stellar correction for either the \madcows\ or SPT subsample
calculated above. The systematic uncertainties inherent in this method
are discussed in \textsection{\ref{Ssec: systematics}}.

A plot of \fstar\ versus \mfh\ is shown in the upper panel of Figure
\ref{Fig: fstarvm500}, in which the red diamonds represent the
infrared-selected \madcows\ clusters and the blue circles represent
the SZ-selected SPT clusters. The dashed green line is the
low-redshift relation found by \citet[][hereafter G13]{GZZ13} and the
black error bars on either side of the plot indicate the systematic
error introduced by the background subtraction. For each cluster in
both samples the stellar mass fraction was calculated without any
stellar rejection and then the average stellar correction for the
appropriate sample, as described in \textsection{\ref{Ssec: LF}}, was
applied in order to achieve a consistent stellar correction for all
the clusters in each sample.

On average, the \madcows\ clusters do not have significantly higher
stellar mass fractions than the SPT clusters. There is a sizable
systematic error, largely from the background subtraction, which
is both larger than the statistical error and mass dependent, but it
should affect both samples to the same degree and thus does not affect
the direct comparison. This is dicussed further in
\textsection{\ref{Ssec: systematics}}. To ensure that this comparison
of \fstar is unrelated to the trend of \fstar\ with mass seen at low
redshift, we also divide out the G13 trend line, as shown in the lower
half of Figure \ref{Fig: fstarvm500}. The errors on the resulting
G13-normalized means for each sample are calculated from bootstrap
resampling and shown as horizontal pink and cyan bars across the
data. This normalization still does not show a significant difference
between the mean of the \Nmadcows\ \madcows\ clusters and the \nSPT\
SPT clusters, though there is still a relatively large error on
the individual \fstar\ errors for both sets of clusters. Stellar
masses and stellar mass fractions for the \madcows\ clusters are given
in Table \ref{Table: fstars}.

\begin{deluxetable}{llccc}
  \tabletypesize{\normalsize} \tablecaption{\madcows\ Stellar Mass Fractions \label{Table: fstars}} \tablewidth{0pt} \tablehead{
    \colhead{ID} & \colhead{$z$} & \colhead{\mfh} &
    \colhead{$M_\star$} & \colhead{\fstar}\\
    \colhead{} & \colhead{} & \colhead{$(10^{14}~\msol)$} &
    \colhead{$(10^{12}~\msol)$} & \colhead{$(10^{-2})$}}
  \startdata
  MOO\ J0037$+$3306 & 1.139 & $2.28^{+0.64}_{-0.61}$ & $4.48\pm0.15$ & $1.97^{+0.56}_{-0.53}$ \\
  MOO\ J0105$+$1323 & 1.143 & $3.92^{+0.46}_{-0.44}$ & $10.73\pm0.19$ & $2.74^{+0.32}_{-0.31}$ \\
  MOO\ J0123$+$2545 & 1.224 & $3.82^{+0.85}_{-0.80}$ & $6.43\pm0.16$ & $1.68^{+0.38}_{-0.35}$ \\
  MOO\ J0319$-$0025 & 1.194 & $3.03^{+0.53}_{-0.46}$ & $2.50\pm0.12$ & $0.82^{+0.15}_{-0.13}$ \\
  MOO\ J1014$+$0038 & 1.229 & $3.22^{+0.36}_{-0.31}$ & $5.44\pm0.16$ & $1.69^{+0.20}_{-0.17}$ \\
  MOO\ J1111$+$1503 & 1.32 & $2.02^{+0.29}_{-0.30}$ & $4.27\pm0.13$ & $2.11^{+0.31}_{-0.32}$ \\
  MOO\ J1142$+$1527 & 1.189 & $5.36^{+0.55}_{-0.50}$ & $7.43\pm0.18$ & $1.39^{+0.15}_{-0.13}$ \\
  MOO\ J1155$+$3901 & 1.009 & $2.53^{+0.50}_{-0.51}$ & $2.60\pm0.11$ & $1.03^{+0.21}_{-0.21}$ \\
  MOO\ J1231$+$6533 & 0.99 & $4.56^{+1.23}_{-0.96}$ & $3.64\pm0.12$ & $0.80^{+0.22}_{-0.17}$ \\
  MOO\ J1514$+$1346 & 1.059 & $1.85^{+0.65}_{-0.77}$ & $6.30\pm0.13$ & $3.40^{+1.20}_{-1.42}$ \\
  MOO\ J1521$+$0452 & 1.312 & $3.59^{+1.02}_{-0.92}$ & $6.77\pm0.17$ & $1.89^{+0.54}_{-0.49}$ \\
  MOO\ J2206$+$0906 & 0.951 & $2.59^{+0.91}_{-0.72}$ & $5.58\pm0.12$ & $2.16^{+0.76}_{-0.60}$ \\
  \enddata
\end{deluxetable}


As the vertical red and blue error bars in the lower panel of Figure
\ref{Fig: fstarvm500} show, the scatter in the SPT stellar mass
fractions is larger than that of the \madcows\ clusters. There is also
a much larger range in the SPT stellar mass fractions, with an order
of magnitude separating the highest \fstar\ clusters from the
lowest. The scatter in \fstar\ seen in the \madcows\ clusters is
lower, but may not be representative of the general cluster population
because of two selection biases. First, \madcows\ is a stellar
mass-selected cluster sample. As such, it may be biased toward systems
with higher-than-average \fstar\ values. Second, this particular
subset of \madcows\ clusters consists of the most significant
detections from the first stage of the study, so may not be
representative of the sample or of clusters as a whole. {\bf We do not
expect the different redshift distributions to introduce a bias,
however, as we find no evidence that \fstar\ evolves with redshift.}
The SPT clusters, however, should provide a fair sample of the mean
value and scatter of the stellar mass fraction at the redshift of
those SZ-selected clusters because they are selected independently of
those components. {\bf We compared the stellar mass fractions of the
\madcows\ and SPT samples using a Kolmogorov-Smirnov test and found
they were consistent with being drawn from the same underlying
distribution.}

\begin{figure*}[bthp]
\center{\includegraphics[width=18cm]{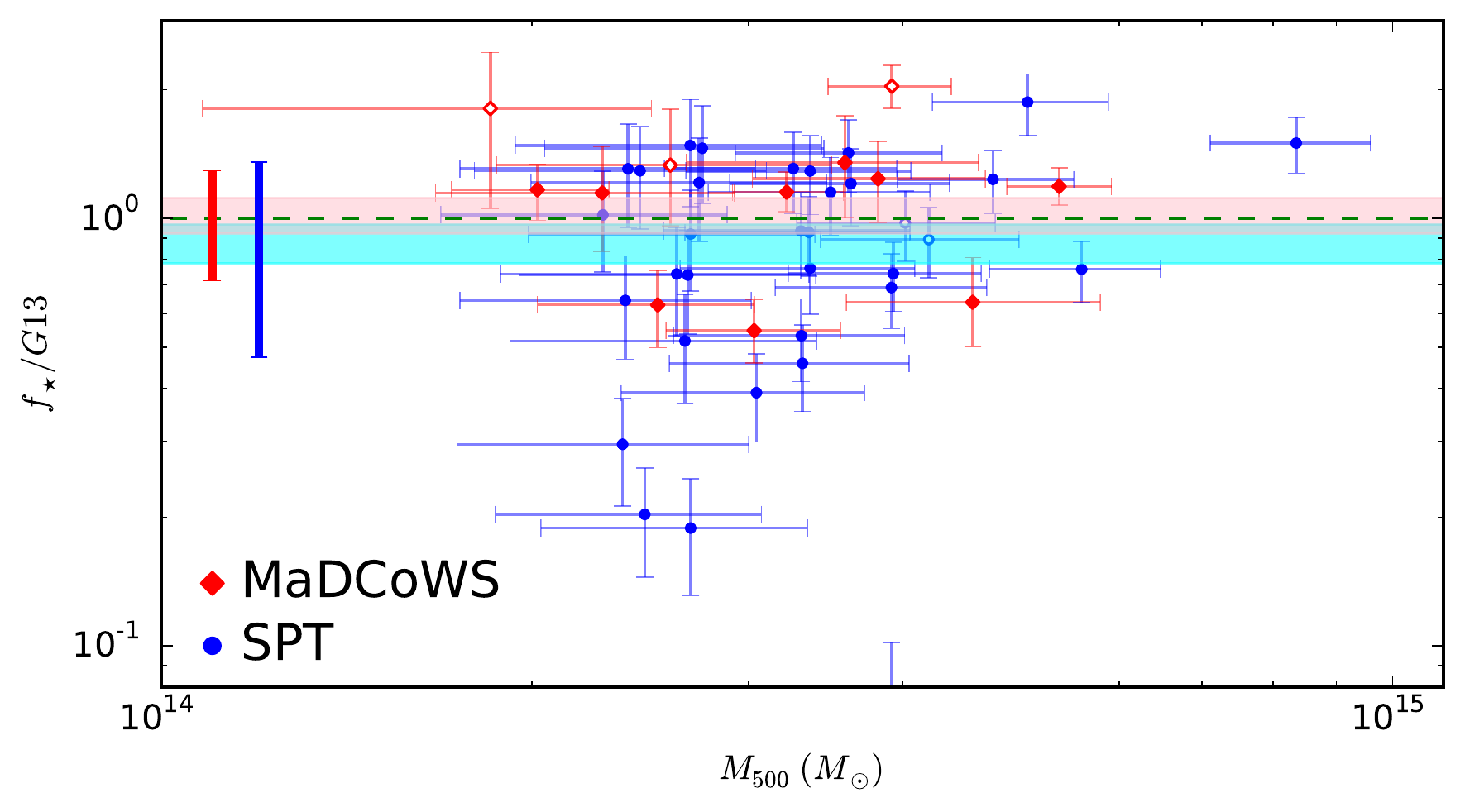}}
\caption{The same as the lower panel of Figure \ref{Fig: fstarvm500},
with the merging \madcows\ clusters (now shown as open diamonds)
removed from the calculation of the mean normalized \fstar. The effect
of removing these clusters for which the total mass is known to be
underestimated relative to the stellar mass is to drop the
G13-normalized mean to \fstar$/G13$ $=1.02\pm0.10$, $1.0\sigma$ higher
than the unchanged SPT mean.}
\label{Fig: fstar w/mergers}
\end{figure*}

The \madcows\ sample contains three clusters known to be merging from
high-resolution {\it Chandra X-ray Observatory} follow-up observations
\citep{Gonzalez+18}.  Previous studies of the effect merging has on
the inferred $Y_{SZ}$ mass of a cluster have produced mixed
conclusions, with some \citep[\eg][]{Poole+07,Krause+12} finding that
major mergers bias the inferred $Y_{SZ}$ mass of a system low for most
of the observed timescale and others, \citep[\eg][]{Marrone+12}
finding the $Y_{SZ}$ mass of merging clusters was overestimated. We do
not expect merging to affect the observed richness of a cluster in the
same way as the mass, however, so any effect on the inferred mass will
bias our measurement of \fstar. We do not have X-ray data for the full
\madcows\ sample or the comparison SPT sample, so we cannot fully
remove mergers from our current analysis. However the effect of
excluding these clusters, for which we know our \fstar\ measurement is
likely to be wrong, is shown in Figure \ref{Fig: fstar w/mergers}. The
clusters are plotted in the same manner as the lower part of Figure
\ref{Fig: fstarvm500}, however the three clusters known to be mergers
are now plotted as open red diamonds and the mean is recalculated to
exclude them. Although they are not large outliers, the three merging
systems do have the highest normalized stellar mass fractions of the
\madcows\ sample. When they are excluded, the mean-normalized \fstar\
for \madcows\ decreases to \fstar$/G13$ $=1.02 \pm 0.10$, still higher
than that of the SPT clusters, but now consistent within 1
$\sigma$. We also removed two clusters from the SPT sample identified
as mergers in \citet[][shown as open circles]{Nurgaliev+17} which did
not affect the mean \fstar$/G13$ of the SPT clusters.

\section{Discussion}
\label{Sec: discussion}
\subsection{Comparison of Stellar Mass Fractions}
\label{Ssec: fstar discussion}
As discussed above, Figures \ref{Fig: fstarvm500} and \ref{Fig: fstar
w/mergers} show that the average stellar mass fraction in the
\madcows\ sample is not significantly higher than that of the SPT
sample, though there is considerable scatter. To confirm that this is
not an artifact of the trend of \fstar\ with mass we normalized all
the \fstar\ measurements relative to the G13 relation and measured the
normalized mean \fstar\ for both samples, shown in the lower panel of
Figure \ref{Fig: fstarvm500}. While the mean normalized \fstar\ for
\madcows, \irnormmean, is higher than the corresponding mean for the
SPT sample, \sznormmean, these are consistent within \normsig.


\subsection{Scatter in the Stellar Mass Fraction}
\label{Ssec: SPT scatter}
The SZ-selected SPT clusters are best-suited to measure the
scatter in \fstar\ at high-redshift as they are selected independently
of stellar content and thus should represent an unbiased
sampling of the stellar mass fraction in the full cluster
population. The large range in \fstar\ seen in this sample,
approximately an order of magnitude (see Figure \ref{Fig:
fstarvm500}), is perhaps surprising. As Figure \ref{Fig: richness
comparison} shows, however, this variation is clearly apparent in a
visual inspection of the richnesses of two clusters with the same halo
mass. Although both clusters in this figure have an SZ mass of \mfh\
$= 2.7\teneft\msol$ \citep{Bleem+15}, SPT-CL\ J0154-4824 (left) has a
stellar mass fraction of \fstar\ $=(2.8 \pm 0.9)\times10^{-3}$ whereas
SPT-CL\ J2148-4843 (right) has a stellar mass fraction of \fstar\
$=(2.6 \pm 0.7)\times10^{-2}$, an order of magnitude higher.

\begin{figure*}[bthp]
\center{\includegraphics[width=18cm]{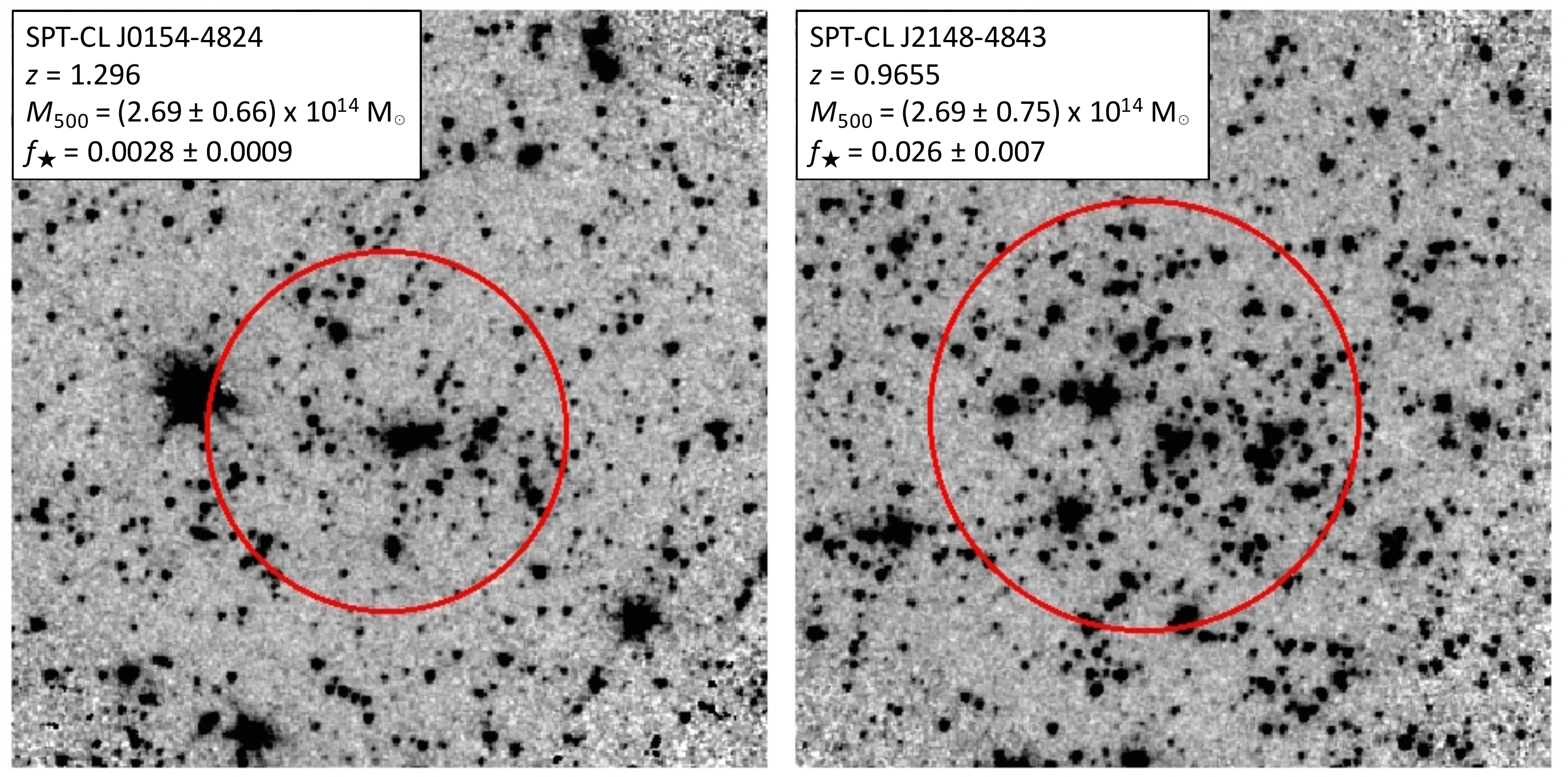}}
\caption{IRAC \cho\ images of SPT-CL\ J0154-4824 (left) and SPT-CL\
J2148-4843 (right) showing the large difference in richness between
clusters of the same halo mass. The projected \rfh\ of each cluster is
shown as a red circle. The difference in the angular size of the two
circles is due to the redshift difference, which boosts the richness
by 28\% in the nearer cluster, but the comparison is relatively
unaffected by the differential K-correction between the clusters as
$\mstar$ in \spitzer\ \cho\ is not significantly different between the
two redshifts.}
\label{Fig: richness comparison}
\end{figure*} 

The \madcows\ clusters in this work do not exhibit the same wide
peak-to-trough range of stellar mass fractions nor as large a scatter,
measured by the standard deviation of \fstar, presumably because they
represent the high-richness end of an infrared-selected sample rather
than a fair cross-section of all clusters. We attempt to quantify the
{\it intrinsic} scatter in \fstar\ of both samples about their
respective means, independent of our measurement errors, by assuming
that the reduced chi-squared will be equal to unity when all the
errors are included in the error budget. We therefore set the reduced
chi-squared for each sample to unity and solve for the intrinsic
scatter term. We find a significant intrinsic scatter, $\sigma_{\ln
f_\star}\sim0.4$ dex for the SPT and $\sigma_{\ln f_\star}\sim0.3$ dex
for \madcows. This discrepancy supports the idea described in
\textsection{\ref{Ssec: fstar}}, that the \madcows\ clusters may not
provide a fair measurement of the scatter in \fstar\ due to their
selection, but the SPT clusters should. By the same token, the
\madcows\ clusters should provide fair measurements of the scatter in
\fgas\ that the SZ-selected surveys may not; this is a topic for
future analyses with \madcows. The SPT clusters show a larger
intrinsic scatter in \fstar\ than is predicted in simulations, such as
those of \citet{Kravtsov+05}, \citet{Ettori+06} and
\citet{Planelles+13}. Very recently, IllustrisTNG \citep{Pillepich+17}
directly measured the scatter in the stellar-total mass relationship
in simulated clusters at $z=0$ and $z\sim1$ and found a very low
scatter in the relationship, only 0.07 dex. Some of the low
values and high scatter in the SPT \fstar\ measurements may be due to
the masses of low signal-to-noise clusters being overestimated. The
clusters we use go to the low signal-to-noise limit of the SPT-SZ
catalog and it is possible that some of these are lower mass clusters
that scattered up above the cutoff. {\bf If we exclude these clusters,
  the intrinsic scatter of the SPT sample becomes consistent with that
  of the \madcows\ clusters.} This effect notwithstanding,
understanding the baryonic processes causing the remaining large
intrinsic scatter in stellar mass fraction, for which the \madcows\
measurement of $\sigma_{\ln f_\star}\sim0.3$ dex may be considered a
lower limit, is a challenge for the next generation of cosmological
simulations.

\subsection{Comparison to Other Works}
\label{Ssec: External comparison}
Given the systematic uncertainties described above, it is difficult to
make direct comparisons to other works with different
systematics. Nevertheless, other works with similar methodologies
provide good external checks on our results, and in particular, allow
us to test the effect of infrared- versus ICM-selection.

\citet{Chiu+18} also measured \fstar\ for 84 clusters from the SPT-SZ
survey, some of which overlap with our SPT comparison clusters. We do
not expect to find the same \fstar\ values for these clusters, as they
use a slightly different cluster mass estimation
\citep[from][]{deHaan+16} and an SED-fitting method to calculate
stellar mass. Nevertheless, their average value for \fstar\ is
consistent with ours for the clusters in the same range of mass and
redshift. 


\citet{Hilton+13} reported stellar and total masses for a sample of 14
SZ-selected clusters from the Atacama Cosmology Telescope (ACT)
in a redshift range of $0.28 \le z \le 1.06$. They have a mean stellar
mass fraction of \fstar\ = $0.023\pm0.003$, which is larger than what
we find for our SZ-selected clusters. However, we use a
\citet{Chabrier03} IMF to calculate stellar mass-to-light ratios which
results in lower stellar masses than the \citet{Salpeter55} IMF
\citet{Hilton+13} used. Accounting for the difference in stellar mass
resulting from the choice of IMFs (0.24 dex), our results are
consistent with theirs.

Similarly, \citet{vanderBurg+14} reported stellar and halo masses for
ten red sequence-selected clusters in a similar redshift range as
ours. Using SED-fitting to determine the stellar mass of each galaxy,
they find a mean stellar mass fraction for their IR-selected clusters
of \fstar\ $=0.013\pm0.002$. This is consistent with our \madcows\
mean of \fstar\ $=$ \irmean, however their method of calculating
stellar mass has different systematics to ours. Correcting for these,
as described below, shifts their average stellar mass fraction higher
than the \madcows\ value, but it remains consistent with the G13 trend
due to their lower mass range. When we divide out the G13 line in the
same manner as in Figure \ref{Fig: fstarvm500}, we find they have an
average normalized stellar mass fraction of \fstar/G13 $= 0.98$,
consistent with what we find for \madcows.

\begin{figure}[bthp]
\includegraphics[width=8.5cm]{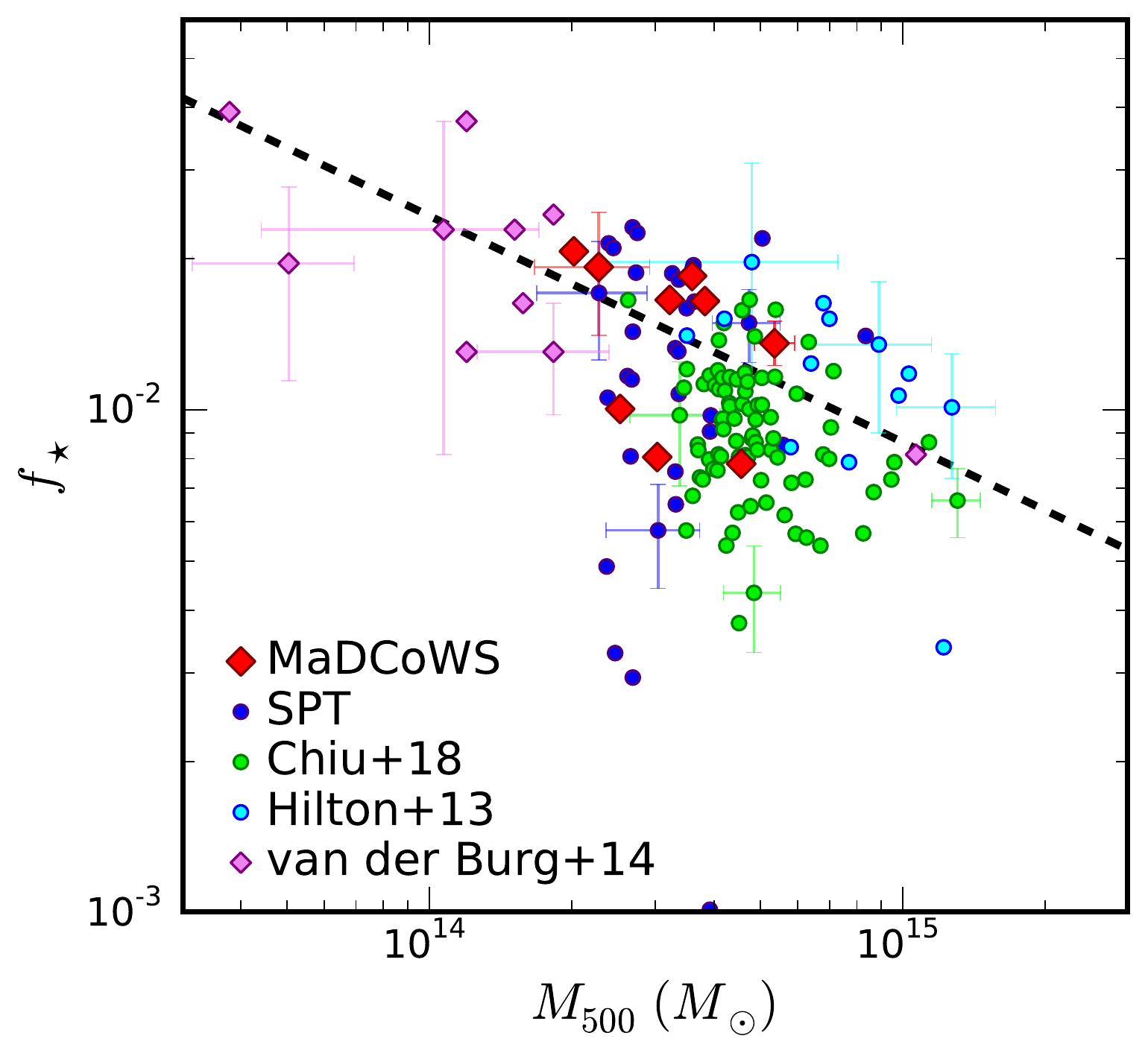}
\caption{Comparison of the \fstar\ measured in this work (red
diamonds, blue circles) to \fstar\ measured by \citet[][green
circles]{Chiu+16}, \citet[][cyan circles]{Hilton+13} and \citet[][violet
diamonds]{vanderBurg+14}. All samples have been adjusted to be
consistent with our methodology. Error bars are plotted for three
representative clusters for each sample.}
\label{Fig: other works comparison}
\end{figure}

Figure \ref{Fig: other works comparison} shows \fstar\ versus \mfh\
for our \madcows\ and SPT clusters plotted alongside the values found
by the studies described above. To make a meaningful comparison, we
corrected the \citet{Hilton+13} and \citet{vanderBurg+14} results to a
Chabrier IMF. We further corrected the latter for the offset between
SED-fitted and $M/L$-based stellar masses reported in that work. The
infrared-selected \madcows\ and \citet{vanderBurg+14} clusters are
plotted as red and violet diamonds, and the SZ-selected
SPT clusters in this work, the \citet{Chiu+18} SPT clusters and the
\citet{Hilton+13} ACT clusters are plotted as blue, green and cyan
circles, respectively. The SZ-selected studies again find
broadly similar stellar mass fractions to the infrared-selected
studies, consistent with what we find here. The G13 relation is
plotted as a dashed line and for each sample error bars are plotted
for three representative clusters.

\subsection{Systematics}
\label{Ssec: systematics}
There are three main sources of systematic error in our analysis. The
largest is due to our background subtraction; this error is
represented by the black error bars in Figure \ref{Fig:
fstarvm500}. We quantify the size of this uncertainty by measuring the
background luminosity from the SDWFS field in $1\arcmin$ radius
cutouts across the field and measure the scatter in this background to
estimate small-scale variation due to clustering. We add this scatter
in quadrature with the field-to-field scatter derived by comparing
SDWFS to similar measurements in the EGS \citep{EGS} and COSMOS fields
\citep{COSMOS}. Since this is an error in the luminosity---and
therefore the stellar mass---of each cluster, the size of the
systematic error in \fstar\ decreases with increasing \mfh. This
systematic error is a uniform shift affecting both the \madcows\ and
SPT clusters equally, so it does not affect our comparison of the
infrared and ICM selection methods.

The second source of systematic uncertainty in the absolute value of
\fstar\ for our clusters is our choice of stellar mass-to-light
ratio. There are two components to this systematic. The first is the
choice of tau model described in \textsection{\ref{Ssec: stellar
mass}}, but this is a small effect. The 1.6\um\ bump is largely
insensitive to the star formation history of the galaxy, so varying
tau does not have a large effect on the M/L ratio. The second
component is the choice of IMF. We use a \citet{Chabrier03} IMF, but
other choices, such as the \citet{Salpeter55} IMF, are also
common. This has a large effect on our M/L ratio, almost doubling it
for a 1 Gyr tau model. However, since this is easily corrected for
and does not affect any comparisons we make, we do not include it in
our systematic error bar in Figure \ref{Fig: fstarvm500}.


A final possible source of systematic uncertainty stems from our
rejection of cluster non-members using magnitude cuts. Our choice of
$\mstar - 2$ as a brightness threshold strikes a balance between
maximizing the bright members included and minimizing the inclusion of
bright foreground interlopers. Although this choice is a somewhat
arbitrary threshold, changing it has only a small effect on our values
for \fstar\ since we already statistically correct for non-member
contamination, and one that is quite consistent from
cluster-to-cluster. It does not make an appreciable difference to our
analysis.

Our faint-end cutoff leads to a modest underestimate of the total
stellar mass. Integrating a luminosity function with $\alpha=-0.8$
beyond $\mstar +1$ suggests we could be missing $\sim 25\%$ of the
stellar mass from fainter galaxies. If we correct our stellar masses
for this, the result is a simple multiplicative increase of all our
\fstar\ values, but by an amount less than both the scatter and the
existing systematic error. Since this offset affects all clusters
equally, it does not affect the scatter in either sample, or our
comparison between the \madcows\ and SPT stellar mass fractions. As a
practical matter, the large uncertainties in $\alpha$ and $\mstar$
make it difficult to accurately quantify the size of this uncertainty,
and thus we choose not to include it in our analysis.

\section{Conclusions}
\label{Sec: conclusion}
We have measured the stellar mass fractions of \Nmadcows\
infrared-selected clusters from \madcows\ and \nSPT\ SZ-selected
clusters from the SPT-SZ survey and found little difference in average
\fstar\ between the two selection methods. We measured \fstar\ using
IRAC \cho\ images of the clusters at $z\sim1$ as a proxy for stellar
mass along with total masses derived from SZ measurements. We found
that when accounting for mergers in the \madcows\ sample and
normalizing over the trend of stellar mass fraction with total mass,
the infrared-selected \madcows\ clusters have an average stellar mass
fraction of \irnormmean, higher than the average stellar
mass fraction of \sznormmean\ for the SPT, but not
significantly so.

We also compare our results to those of \citet{Hilton+13},
\citet{vanderBurg+14} and \citet{Chiu+16} who also looked at stellar
mass fractions in cluster samples of comparable mass and redshift to
ours. When we correct for the differences between our methodologies
and those of the other studies, we find our results are consistent
with all three and they support our result that infrared-selected
clusters do not have an appreciably higher mean \fstar than
SZ-selected clusters. We also compare the value we calculate for
$\mstar$ of the IRAC \cho\ luminosity function to that found by
\citet{Muzzin+08}, \citet{Mancone+10} and \citet{Wylezalek+14} and
find similar results.

We found an unexpectedly large range in the stellar mass fractions of
individual clusters in the SPT sample and a larger range and scatter
in \fstar\ than in our \madcows\ clusters. It is possible that the
SZ-selected SPT clusters give a fairer sample of the full range of
\fstar\ than the infrared-selected \madcows\ clusters do. Future work
with \madcows\ will compare \fgas\ measurements in infrared- and
SZ-selected cluster samples to look for a comparable selection effect
in the latter.

Finally, we have presented SZ observations of \newSZ\ new \madcows\
clusters and new spectroscopic redshifts for five clusters. Among the
SZ observations of the \newSZ\ new \madcows\ clusters is MOO\
J1521+0452, which at $z=1.31$ is one of the most massive clusters yet
found at $z\ge1.3$. Along with the previous discovery of a cluster of
\mfh\ $= (5.36^{+0.55}_{-0.50})\times10^{14}~\msol$ at $z=1.19$,
reported in \citet{Gonzalez+15}, this further demonstrates the ability
of \madcows' nearly all-sky infrared selection to find the most
massive clusters at high redshifts.

\acknowledgments 
Support for CARMA construction was derived from the Gordon and Betty
Moore Foundation, the Kenneth T. and Eileen L. Norris Foundation, the
James S. McDonnell Foundation, the Associates of the California
Institute of Technology, the University of Chicago, the states of
California, Illinois, and Maryland, and the National Science
Foundation. CARMA development and operations are supported by the
National Science Foundation under a cooperative agreement, and by the
CARMA partner universities; the work at Chicago was supported by NSF
grant AST- 1140019. Additional support was provided by PHY- 0114422.
This work is based in part on observations made with the {\it Spitzer
Space Telescope}, which is operated by the Jet Propulsion Laboratory,
California Institute of Technology under a contract with NASA. {\bf Support for this work was
  provided by NASA through an award issued by JPL/Caltech.} This
publication makes use of data products from the {\it Wide-field
Infrared Survey Explorer}, which is a joint project of the University
of California, Los Angeles, and the Jet Propulsion
Laboratory/California Institute of Technology, funded by the National
Aeronautics and Space Administration. Some of the data presented
herein were obtained at the W.M. Keck Observatory, which is operated
as a scientific partnership among the California Institute of
Technology, the University of California and the National Aeronautics
and Space Administration. The Observatory was made possible by the
generous financial support of the W.M. Keck Foundation. This work was
based in part on observations obtained at the Gemini Observatory,
which is operated by the Association of Universities for Research in
Astronomy, Inc., under a cooperative agreement with the NSF on behalf
of the Gemini partnership: the National Science Foundation (United
States), the National Research Council (Canada), CONICYT (Chile),
Ministerio de Ciencia, Tecnolog\'{i}a e Innovaci\'{o}n Productiva
(Argentina), and Minist\'{e}rio da Ci\^{e}ncia, Tecnologia e
Inova\c{c}\~{a}o (Brazil). This work has made use of data from the
European Space Agency (ESA) mission {\it Gaia}
(\url{https://www.cosmos.esa.int/gaia}), processed by the {\it Gaia}
Data Processing and Analysis Consortium (DPAC,
\url{https://www.cosmos.esa.int/web/gaia/dpac/consortium}). Funding
for the DPAC has been provided by national institutions, in particular
the institutions participating in the {\it Gaia} Multilateral
Agreement. {\bf Funding for this program is provided by NASA through
the NASA Astrophysical Data Analysis Program, award
NNX12AE15GB. Support for this work was provided by the National
Aeronautics and Space Administration through Chandra Award Number
GO6-17130X issued by the Chandra X-ray Center, which is operated by
the Smithsonian Astrophysical Observatory for and on behalf of the
National Aeronautics Space Administration under contract
NAS8-03060.}

\bibliographystyle{aasjournal}
\bibliography{madcows}

\end{document}